\documentclass[aps, pra, reprint, longbibliography,
superscriptaddress]{revtex4-1}

\usepackage{hyperref}
\usepackage{graphicx}
\usepackage[utf8]{inputenc}
\usepackage{xcolor}
\usepackage{amsmath, amssymb}
\usepackage{physics}
\usepackage{ulem}

\begin{document}

\preprint{APS/123-QED}

\title{Tilted dipolar bosons in the quasi-2D regime: from liquid stripes to droplets.}

\author{J. S\'anchez-Baena}
\email{juan.sanchez.baena@upc.edu}
\affiliation{Departament de F\'isica, Universitat Polit\`ecnica de Catalunya, Campus Nord B4-B5, 08034 Barcelona, Spain}

\date{\today}

\begin{abstract}

We characterize a system of tilted dipoles in a quasi two-dimensional (flattened) geometry and in the thermodynamic limit. We consider a finite trapping in the $z$-axis achievable in current experiments. We compute the phase diagram of the system at its equilibrium density for high tilting angles, where it becomes self-bound, and a striped liquid state emerges.
To characterize the system, we perform a variational calculation, which is benchmarked with the solution of the extended Gross-Pitaevskii equation. We connect the phenomenology in the thermodynamic limit to the physics of the finite-size system, provide parameters for the realization of potentially supersolid striped states and study the critical number for dipolar droplet formation.
Our results are helpful to guide potential experiments in the study of dipolar atoms in quasi two-dimensional geometries in the dipole-dominated regime.

\end{abstract}

\maketitle

\section{\label{sec:introduction}Introduction}

Ultracold quantum gases with dipolar interactions have risen to prominence in the last decade as a platform to study a plethora of exotic phenomena in a controlled environment. The interplay between the anisotropy and long-range behaviour of the dipole-dipole interaction (DDI) is responsible for the emergence of dipolar droplets~\cite{Pfau:nature:2016,Pfau:nature2:2016,Pfau:PRL:2016,ferlaino16,
bottcher19}, which constitute an ultra-dilute liquid state, supersolids~\cite{Modugno:PRL:2019,Pfau:PRX:2019,Ferlaino:PRX:2019,
Tanzi:Nature:2019,Guo:Nature:2019,Tanzi:Science:2021,norcia21:nature,
BiagioniPRX2022,Ferlaino:PRL:2021}, which combine two apparently mutually exclusive properties like global phase coherence and spatial periodicity, or anomalous thermal behaviour~\cite{Ferlaino:PRL:2021,baena22,baena24,He2024}, in the form of the promotion of supersolidity by increasing the temperature. Recently, many efforts are being directed towards the realization of ultracold atom experiments in a quasi two-dimensional geometry, in a regime where physical phenomena exclusive to the two-dimensional realm can be accessed. This includes the measurement of the Berezinskii-Kosterlitz-Thouless (BKT) transition for a system of $^{87}$Rb atoms~\cite{sunami2022:prl}, the realization of a Bose glass in a quasicrystalline optical lattice~\cite{Yu2024:science}, or the direct probing of a 1D-2D crossover for caesium atoms~\cite{Guo2024:science}, to name a few. In the case of dipolar systems, quasi two-dimensional systems have the added interest of studying the interplay between the unique properties of the DDI and a dimensional constrain. Due to the anisotropy of the DDI, limiting the motion of particles in one direction introduces a new variable with respect to the three-dimensional case: the polarization direction of the dipoles. Moreover, quantum fluctuations, which play a crucial role in dipolar systems by arresting the mean-field collapse~\cite{Lima:2011eq,pelster12,Wachtler:2016kk} depend significantly on dimensionality~\cite{edler_PRL_2016,Zin_2021}. Because of this, experimental activity targeting these systems has emerged recently. In particular, a bilayer of Dy atoms with a layer width on the sub-$50$ nm scale has been realized~\cite{li_du2024:science}, and the BKT transition has been experimentally accessed for a dipolar system, where the wave function of atoms along the tightly trapped direction approximately corresponds that of the harmonic oscillator ground state~\cite{he2024:arxiv}. In light of this experimental activity, it is necessary to provide theoretical calculations pointing at the relevant parameters for the observation of novel physical phenomena exclusive to the quasi two-dimensional regime.

Several theoretical works have been produced along the years with the goal of characterizing dipolar systems in flattened geometries under different specific conditions. These include the study of the ground state, collapse instability and excitations of the gas phase at the mean field level~\cite{fischer_06_PRA, boudjemaa_2013_PRA, mishra_2016_PRA, Baillie_2015_NJP}, including finite temperature~\cite{pengtao2021}, and the characterization of the stripe phase for bosons~\cite{fedorov_2014_PRA,mishra_2016_PRA,aleksandrova_2024_PRA}, fermions~\cite{marchetti2013:prb,block2014:prb} and mixtures~\cite{lee_2024_PRA}, where a spin-stripe phase emerges. The system has also been studied in the 3D limit~\cite{zhang19, zhang21, maucher24, cinti2025} and in the presence of a two-dimensional lattice, both at zero and finite temperature~\cite{amrey2015,soumik:PRA:2019}. In particular, the results in the 3D limit show the importance to modify the tilting angle of the dipoles to reduce the necessary density for the experimental realization of the stripe phase, which is excessively large in the case of dipoles perpendicular to the plane of motion, and thus induces important three-body losses~\cite{zhang21}. Through this approach, striped arrangements of tilted dipoles have been observed in experiments~\cite{wenzel_2017:PRA}, with the number of stripes being controllable with the trap aspect ratio. In the strict two-dimensional case, quantum Monte Carlo methods have been employed to study the phases of the system for the case of a single layer~\cite{macia12b,macia12,macia14,bombin17:PRL,bombin19:PRA}, a bilayer~\cite{guijarro22:PRL} and a multilayer~\cite{guijarro2024} configuration.

\begin{figure*}[t]
\centering
\includegraphics[width=\linewidth]{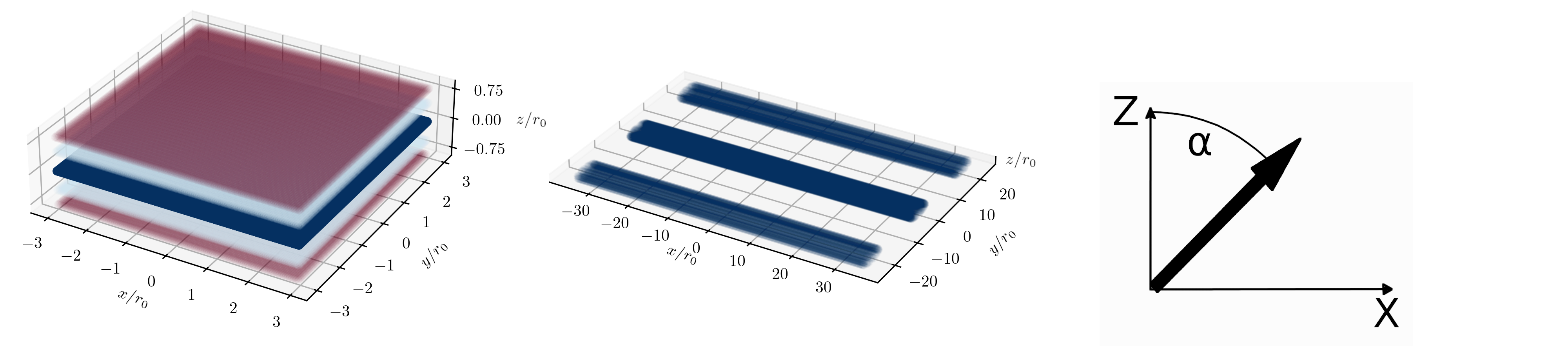}
\caption{Schematic representation of the system of dipoles in a flattened geometry. Left: isodensity surfaces in the homogeneous liquid phase. The different colors correspond to surfaces where $n = 0.5 n_{\rm peak}$ (red), $n = 0.8 n_{\rm peak}$ (light blue) and $n = n_{\rm peak}$ (dark blue), with $n$ the 3D density and $n_{\rm peak}$ the peak density. Center: isodensity surface ($n = 0.5 n_{\rm peak}$ (dark blue)) of the system in the liquid stripe phase.  Right: graphical representation of a polarized dipole (thick black arrow) with tilting angle $\alpha$. }
\label{fig0}
\end{figure*}

Recently, a theoretical work brought the attention into the liquid properties of the system of tilted dipoles in the fully two-dimensional case~\cite{staudinger23:PRA}, where the emergence of an exotic, striped liquid phase in the thermodynamic limit is shown. Nevertheless, the strict two-dimensional limit, which is achieved when $l \ll a$, with $l$ the oscillator length in the z-axis and $a$ the scattering length, lies away from the state of the art experiments, which typically fulfill $l > a$. This has important consequences on relating the phenomenology of the system with the $s$-wave scattering length, which is of crucial importance in experiments, while also impacting greatly the critical number for the formation of droplets, as we shall illustrate below.

In this work, we perform a variational calculation to obtain the phase diagram of an infinite system of tilted dipoles in a flattened geometry at the equilibrium density. In specific cases, we benchmark our variational results with the solution of the extended Gross-Pitaevskii equation (eGPE), which we also employ to perform calculations of a finite size system under a box trap. This paper is organized as follows. In Sec.~\ref{sec:theory}, we present the variational theory and the eGPE. In Sec.~\ref{sec:results}, we present our results for both the thermodynamic limit (Sec.~\ref{sec:bulk}) and the finite size system (Sec.~\ref{sec:finite_size}) while in Sec.~\ref{sec:conclusions} we summarize the main conclusions of our work.

\section{\label{sec:theory}Theory}

We study a system governed by the energy functional

\begin{align}
 &E [\psi] = \int d{\bf r} \psi^{*}({\bf r}) \left[ -\frac{\hbar^2 \nabla^2}{2 m} + U({\bf r}) \right] \psi({\bf r}) \nonumber \\
 &+ \frac{1}{2} \int \int d{\bf r} d{\bf r'} \abs{\psi({\bf r'})}^2 V({\bf r} - {\bf r'}) \abs{\psi({\bf r'})}^2 \nonumber \\
 &+ \int d{\bf r} \epsilon_{\rm BMF} \left( \abs{\psi({\bf r})}^2 \right)
 \label{energy}
\end{align}
where $U({\bf r}) = \frac{1}{2} m \omega^2 z^2$ for the infinite system and $U({\bf r}) = \frac{1}{2} m \omega^2 z^2 + U_{\rm box}(x,y)$ for the finite one with $U_{\rm box}(x,y)$ a box trap. $V({\bf r} - {\bf r'})$ is the contact plus dipole-dipole interaction, i.e.
\begin{align}
 V({\bf r} - {\bf r'}) = \frac{4 \pi \hbar^2 a}{m} \delta\left( {\bf r} - {\bf r'} \right) + \frac{C_{\rm dd}}{4 \pi} \frac{ \left( 1 - 3 \cos^2 \theta_{\alpha} \right) }{\abs{{\bf r} - {\bf r'}}^3} \ , \label{pseudopot}
\end{align}
with $a$ the s-wave scattering length and $\theta_{\alpha}$ the angle between the vector ${\bf r} - {\bf r'}$ and the polarization axis of the dipoles. In the calculations, we change this direction within the $x$-$z$ plane, and define $\alpha$ as the angle between the polarization direction and the $z$-axis (see Fig.~\ref{fig0}). Here $C_{\rm dd} = \mu_0 \mu^2$ for magnetic dipoles. We restrict ourselves to the regime where the collisions between particles can be treated as a three-dimensional process, meaning that $l > a$ with $l = \sqrt{\frac{\hbar}{m \omega}}$, which is necessary for the model interaction of Eq.~\ref{pseudopot} to be valid. We define characteristic energy and length scales given by $r_0 = 12 \pi a_{\rm dd}$ and $E_0 = \hbar^2 / (m r_0^2)$ and present our results in these units. Here, $a_{\rm dd} = m C_{\rm dd}/(12 \pi \hbar^2)$ is the dipole length.

The quantity $\epsilon_{\rm BMF} \left( n = \abs{\psi({\bf r})}^2 \right)$ is an energy functional accounting for beyond mean-field (BMF) effects. In order to compute it, we follow Ref.~\cite{Zin_2021}, where the BMF energy density is obtained for an homogeneous system with periodic boundary conditions on the $z$-axis, thus accounting approximately for the discretized excitations due to finite trapping effects. Remarkably, the scaling law of $\epsilon_{\rm BMF}$ with the density depends on the density regime. This is because at lower densities only the lowest excitations along the $z$-axis are populated (meaning that the system is in the quasi two-dimensional limit) while higher densities yield the 3D scaling. In practice, if we denote by $n_{\rm lim}$ the limiting value of the density where the scaling changes, the BMF energy density can be fitted to~\cite{Zin_2021}
\begin{align}
 \epsilon^{\rm low}_{\rm BMF}(n) &= C_1 n^2 + C_2 n^3 + C_3 n^{5/2} & n < n_{\rm lim} \label{fluc_1} \\
 \epsilon^{\rm high}_{\rm BMF}(n) &= C_4 n^{5/2} & n > n_{\rm lim} \label{fluc_2}
\end{align}
The term $C_3 n^{5/2}$ is included in Eq.~\ref{fluc_1} with respect to the result of Ref.~\cite{Zin_2021} to improve the accuracy of the fit, while Eq.~\ref{fluc_2} shows the well-known density scaling in the 3D limit~\cite{Lima:2011eq, pelster12}.
In the calculation for $\epsilon_{\rm BMF}$, we set the value of the length of the $z$-axis to $L_z = 4 l$, with $l$ the harmonic oscillator length of the trapping, and find $n_{\rm lim}$ numerically (see the Appendix~\ref{sec:dens_lim}). This density is an important parameter, as it helps to determine whether the system is effectively two or three-dimensional.
When evaluating Eq.~\ref{energy}, we have to choose one of the two density functionals from Eqs.~\ref{fluc_1} and~\ref{fluc_2}. The choice is determined by the peak density of the converged ground state. We employ Eq.~\ref{fluc_1} if $n_{\rm peak} \lesssim n_{\rm lim}$ and Eq.~\ref{fluc_2} if $n_{\rm peak} \gg n_{\rm lim}$, with $n_{\rm peak}$ the peak (or maximum) 3D density. It can be seen numerically that the energy functional of Eq.~\ref{fluc_2} lays quantitatively very close (less than a few $\%$ away) to the standard LHY energy density correction in three dimensions given in Refs.~\cite{Lima:2011eq, pelster12}.

At high enough values of the tilting and for a tight confinement along the $z$-axis, the system is expected to host either an unmodulated state, or a striped state as its ground state~\cite{staudinger23:PRA,aleksandrova_2024_PRA}.
Even if a triangular state or a honeycomb state can appear as the ground state of a system of dipoles in a flattened geometry for a given range of densities and scattering lengths~\cite{zhang19,zhang21,cinti2025}, we argue in the Appendix~\ref{sec:triangular}) that this is not the case for the parameters considered in our work.
Thus, in order to find the lowest energy state, we perform a variational minimization of Eq.~\ref{energy} for two different ansatz. These are the senoidal ansatz
\begin{align}
 \psi_{\rm S}({\bf r}) = \sqrt{\frac{n_{\rm 2D}}{\sqrt{\pi} \sigma_z}} \sqrt{1 + A \cos(k y)} e^{ - \frac{1}{2} \left( \frac{z}{\sigma_z} \right)^2 } \ , \label{senoidal}
\end{align}
with $A$, $k$ and $\sigma_z$ variational parameters, and the gaussian ansatz
\begin{align}
 \psi_{\rm G}({\bf r}) &= \sum_{j= -\infty}^{\infty} \int dy' \sqrt{\frac{n_{\rm 2D} L_y}{\pi \sigma_z \sigma_y}} e^{ - \frac{1}{2} \left[ \left( \frac{y'}{\sigma_y} \right)^2 + \left( \frac{z}{\sigma_z} \right)^2 \right] } \nonumber \\
 &\times \delta(y + jL_y - y') \ , \label{gaussian}
\end{align}
where the variational parameters are $L_y$, $\sigma_y$ and $\sigma_z$. In both expressions, $L_y$ is the length of the unit cell in the $y$-axis ($L_y = 2 \pi / k$ for the senoidal ansatz), $n_{\rm 2D} = \frac{N}{L_x L_y}$ is the 2D density and the normalization ensures that $N = \int_{-L_x/2}^{L_x/2} \int_{-L_y/2}^{L_y/2} \int_{-\infty}^{\infty} d{\bf r} \abs{\psi_{\rm S, G}({\bf r})}^2$. The senoidal ansatz converges to an unmodulated state in the limit $A \rightarrow 0$, meaning that with these wave functions we can account for an unmodulated state, a phase coherent, striped state and a highly incoherent striped state.

We insert Eqs.~\ref{senoidal} and~\ref{gaussian} into Eq.~\ref{energy} to obtain expressions for the energy of the system. In a general way, we split the contributions of the energy into
\begin{align}
 E_{\rm S, G} &= E_1 + E_{\rm V} + E_{\rm BMF} \label{e_split} \ , \\
 E_1 &= \int d{\bf r} \psi^{*}({\bf r}) \left[ -\frac{\hbar^2 \nabla^2}{2 m} + U({\bf r}) \right] \psi_{\rm S, G}({\bf r}) \ , \\
 E_{\rm V} &= \frac{1}{2} \int \int d{\bf r} d{\bf r'} \abs{\psi_{\rm S, G}({\bf r'})}^2 V({\bf r} - {\bf r'}) \abs{\psi_{\rm S, G}({\bf r'})}^2 \ , \\
 E_{\rm BMF} &= \int d{\bf r} \epsilon_{\rm BMF} \left( \abs{\psi_{\rm S, G}({\bf r})}^2 \right) \ ,
\end{align}
where the subindex ``S, G'' of Eq.~\ref{e_split} refers to the senoidal and gaussian ansatz, respectively. The contributions to the energy per particle $E/N$ for the senoidal ansatz are given by
\begin{widetext}
\begin{align}
 E_1/N &= \frac{\hbar^2 A^2 k^2}{8 m} \frac{1}{2 \pi} \int_{-\pi}^{\pi} d \xi \frac{\sin^2(\xi)}{1 + A \cos \xi} + \frac{\hbar^2}{4 m \sigma_z^2} + \frac{m \omega^2 \sigma_z^2}{4} \ ,
 \\
 E_{\rm V}/N &= \frac{n_{\rm 2D}}{2} \left( \tilde{V}(0,0) + \frac{A^2}{2} \tilde{V}(0,k) \right) \ , \label{E_V_senoid}
 \\
 E^{\rm high}_{\rm BMF}/N &= C_4 n^{3/2}_{\rm 2D} \sqrt{\frac{2}{5 \sigma_z^3 \pi^{3/2}}} \frac{1}{2 \pi} \int_{-\pi}^{\pi} d \xi \left( 1 + A \cos \xi \right)^{5/2} \ ,
 \\
 E^{\rm low}_{\rm BMF}/N &= C_1 \frac{1}{\sqrt{2 \pi} \sigma_z} n_{\rm 2D} \frac{1}{2 \pi} \int_{-\pi}^{\pi} d \xi \left( 1 + A \cos \xi \right)^{2} + C_2 \frac{1}{\sqrt{3} \pi \sigma_z^2} n^2_{\rm 2D} \frac{1}{2 \pi} \int_{-\pi}^{\pi} d \xi \left( 1 + A \cos \xi \right)^{3}  \nonumber \\
 &+ C_3 n^{3/2}_{\rm 2D} \sqrt{\frac{2}{5 \sigma_z^3 \pi^{3/2}}} \frac{1}{2 \pi} \int_{-\pi}^{\pi} d \xi \left( 1 + A \cos \xi \right)^{5/2} \ ,
\end{align}
\end{widetext}
where the term $\tilde{V}(k_x,k_y)$ of Eq.~\ref{E_V_senoid} is~\cite{fedorov_2014_PRA}
\small
\begin{align}
 \tilde{V}(k_x,k_y) &= \frac{2 \sqrt{2 \pi} \hbar^2 a}{m \sigma_z} - \frac{2 \sqrt{2 \pi} \hbar^2 a_{\rm dd}}{m \sigma_z} \nonumber \\
 &+ \frac{6 \hbar^2 a_{\rm dd}}{m} \int_{-\infty}^{\infty} dk_z \frac{k_x^2}{k^2} \exp \left( - \frac{k_z^2 \sigma_z^2}{2} \right) \sin^2 \alpha \nonumber \\
 &+ \frac{6 \hbar^2 a_{\rm dd}}{m} \int_{-\infty}^{\infty} dk_z \frac{k_z^2}{k^2} \exp \left( - \frac{k_z^2 \sigma_z^2}{2} \right) \cos^2 \alpha \ .
\end{align}
\normalsize
 In much the same way, the contributions of $E/N$ for the gaussian ansatz are
\begin{widetext}
\begin{align}
 E_1/N &= \frac{\hbar^2}{2 m} \left( \frac{1}{2 \sigma_y^2} + \frac{1}{2 \sigma_z^2} \right) + \frac{m \omega^2 \sigma_z^2}{4} \ ,
 \\
 E_{\rm V}/N &= \frac{n_{\rm 2D}}{2} \left( \frac{L_y}{\sqrt{\pi}
  \sigma_y} \right)^2 \left( A_0^2 \tilde{V}(0,0) + \sum_{j=1}^{\infty} \frac{A_j^2}{2} \tilde{V}(0,jk) \right) \ , \label{E_V_gaussian}
 \\
 E^{\rm high}_{\rm BMF}/N &= \frac{2 C_4}{5} \left( \frac{n_{\rm 2D} L_y }{\pi \sigma_y \sigma_z} \right)^{3/2} \ ,
 \\
 E^{\rm low}_{\rm BMF}/N &= \frac{C_1}{2} \frac{n_{\rm 2D} L_y}{\pi \sigma_y \sigma_z} + \frac{C_2}{3} \left( \frac{n_{\rm 2D} L_y}{\pi \sigma_y \sigma_z} \right)^2 + \frac{2 C_3}{5} \left( \frac{n_{\rm 2D} L_y }{\pi \sigma_y \sigma_z} \right)^{3/2} \ ,
\end{align}
\end{widetext}
where we have assumed that $L_y \gg \sigma_y$. The coefficients $A_j$ from Eq.~\ref{E_V_gaussian} are obtained from the Fourier series of the density and given by
\begin{align}
 A_j &= f_j \frac{\sqrt{\pi} \sigma_y}{L_y} \exp \left( -\frac{(2 \pi j \sigma_y)^2}{4 L_y^2} \right) \ ,
\end{align}
where $f_0 = 1$ and $f_j = 2$ for $j>0$. The variational minimization of the energy per particle for each ansatz is performed numerically via a Simulated Annealing algorithm~\cite{delahaye:hal-01887543}. In practice, the infinite sum of Eq.~\ref{E_V_gaussian} is truncated in the numerical calculation, and the number of coefficients is chosen to achieve convergent results. We have found it sufficient to sum up to $j = 2000$ to ensure convergence.

Performing a functional minimization of Eq.~\ref{energy} with respect to $\psi^{*}({\bf r})$ with the constrain $N = \int d{\bf r} \abs{\psi({\bf r})}^2$ we obtain the zero temperature, extended Gross-Pitaevskii equation (eGPE), which is given by
\begin{align}
 &\mu \psi({\bf r}) = \left[ -\frac{\hbar^2 \nabla^2}{2 m} + U({\bf r}) + \int d{\bf r'} V({\bf r} - {\bf r'}) \abs{\psi({\bf r'})}^2 \right. \nonumber \\
 &\left. + \eval{\pdv{\epsilon_{\rm BMF}}{n}}_{n=\abs{\psi({\bf r})}^2} \right] \psi({\bf r}) \ .
 \label{eGPE}
\end{align}
We use the eGPE to benchmark our variational calculations in the thermodynamic limit and to study the finite size system under a box trap in the $x$-$y$ plane.

\begin{figure}[t]
\centering
\includegraphics[width=0.8\linewidth]{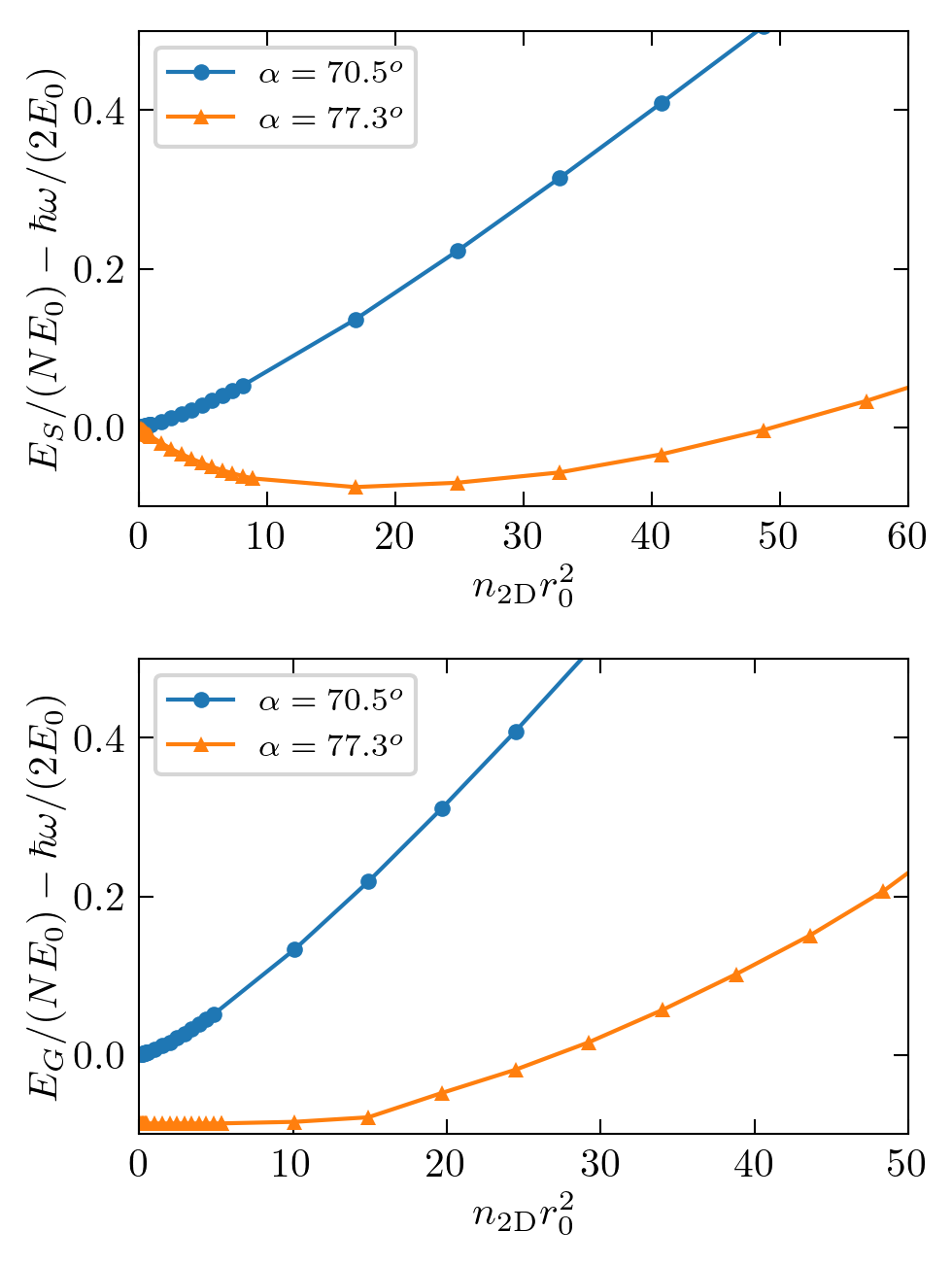}
\caption{Energy per particle for the senoidal (top, see Eq.~\ref{senoidal}) and gaussian (bottom, see Eq.~\ref{gaussian}) ansatz, respectively. The value of the scattering length is set to $a/a_{\rm dd} = 0.7$. We have subtracted the energy of the non-interacting system, $E_{\rm NI}/(N E_0) = \hbar \omega/(2 E_0)$.}
\label{fig2}
\end{figure}

\section{\label{sec:results}Results}

\subsection{\label{sec:bulk}Infinite system}

We start by discussing the results for the infinite system. In this case, the trapping potential in Eq.~\ref{energy} is set to $U({\bf r}) = \frac{1}{2} m \omega^2 z^2$, while the system extends indefinitely in the $x$-$y$ plane. In practice, we apply periodic boundary conditions in the $x$-$y$ plane, with the simulation box determined by the lengths $L_x = L_y$ that minimize the energy per particle at each density. Therefore, by considering infinite replicas of our simulation box, we are taking the limit $N \rightarrow \infty$, $L_y \rightarrow \infty$ for a given 2D density $n_{\rm 2D}$ (thermodynamic limit). Unless specified otherwise, we set the trapping strength along the $z$-axis to $\omega = 2 \pi \times 906$ Hz in all calculations, which in our units corresponds to $\hbar \omega / E_0 = 1$ for $^{164}$Dy atoms. This trapping strength is close to the one employed in the recent experiment~\cite{he2024:arxiv}, where the BKT transition was probed for a dipolar system. For a given value of the $s$-wave scattering length $a$, tilting $\alpha$ and 2D density $n_{\rm 2D}$, we can minimize the energy per particle $E/N$ for each of the two ansatz of Eqs.~\ref{senoidal} and~\ref{gaussian}. In particular, we are interested in the state of the system at the equilibrium 2D density $n^{\rm eq.}_{\rm 2D}$, i.e. the 2D density at which the energy per particle is minimized. We illustrate the different qualitative behaviour of the energy per particle as a function of $n_{\rm 2D}$ for the two ansatz in Fig.~\ref{fig2} for two different sets of parameters $\{a/a_{\rm dd},\alpha\} = \{0.7, 70.47^o\}$ and $\{0.7, 77.35^o\}$. From the figure, we see that $E_{\rm G}/N$ increases monotonously with the density in both cases and the minimum value is achieved for $n_{\rm 2D} \rightarrow 0$, with $N\text{, }L_x L_y \rightarrow \infty$. This is because the gaussian ansatz describes an array of incoherent stripes, spaced by a distance $L_y$. Since the inter-stripe energy is globally repulsive, the minimum energy is achieved for infinitely spaced stripes, characterized by a given ratio of particles per length along the $x$-axis, $N/L_x$. Moreover, in the range of values of $L_y$ where the inter-stripe interactions are neglectable, $E_{\rm G}/N$ stays essentially unchanged for $n_{\rm 2D} L_y = \text{constant}$. The gaussian ansatz describes a striped system that can correspond to either a gas ($E_{\rm G}/N > E_{\rm NI}/N$) or a liquid ($E_{\rm G}/N < E_{\rm NI}/N$), where $E_{\rm NI}/(N E_0) = \hbar \omega/(2 E_0)$ is the energy per particle of the non-interacting system. This is because if $E_{\rm G}/N > E_{\rm NI}/N$, a finite system expands indefinitely to reduce its energy while if $E_{\rm G}/N < E_{\rm NI}/N$, the system prefers to bunch into stripes to do so. Moving on to the results for the senoidal ansatz, $E_{\rm S}/N$ shows two different qualitative trends for the parameters chosen: either it increases monotonously with $n_{\rm 2D}$ or it shows a minimum at a given 2D density. As known from thermodynamics, the former trend corresponds to a gas (expanding) state, while the latter corresponds to a liquid.

\begin{figure}[t]
\centering
\includegraphics[width=\linewidth]{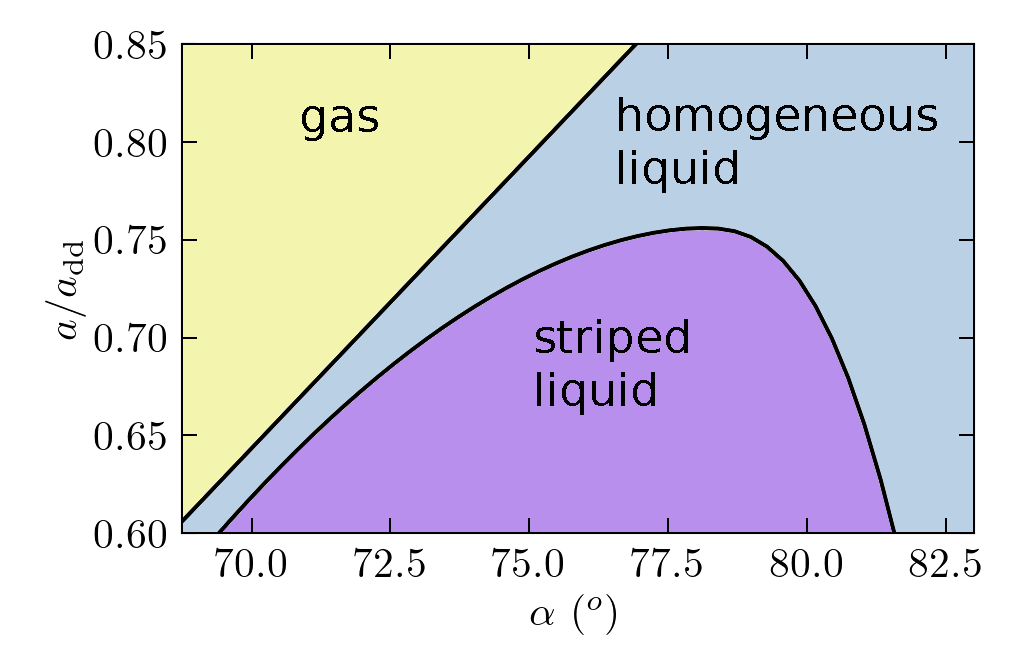}
\caption{Phase diagram of the infinite dipolar system in a flattened geometry at the equilibrium 2D density.}
\label{fig3}
\end{figure}

\begin{figure}[t]
\centering
\includegraphics[width=0.8\linewidth]{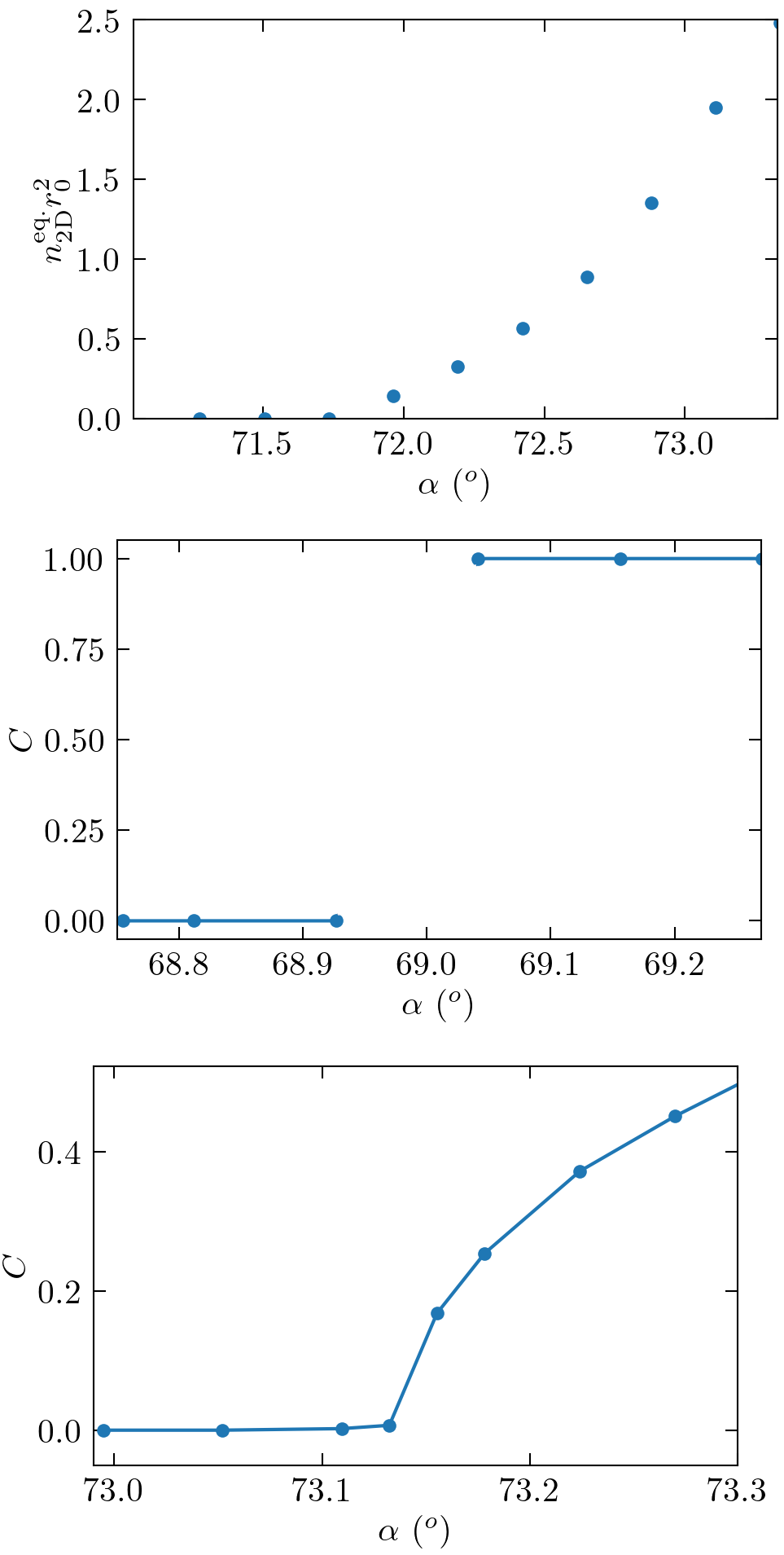}
\caption{2D equilibrium density (top) and contrast of the density ($C = \frac{n_{\rm max.} - n_{\rm min.}}{n_{\rm max.} + n_{\rm min.}}$) (center, bottom) as a function of the tilting angle. The scattering length is set to $a/a_{\rm dd} = 0.7$ in the top and bottom figures while it is set to $a/a_{\rm dd} = 0.6$ in the center figure. }
\label{fig4}
\end{figure}

For each set of parameters $\{a,\alpha\}$ we perform the variational energy minimization for the senoidal and gaussian ansatz and look for the lowest energy state at the equilibrium 2D density. From these calculations, we obtain the phase diagram of the system, which is shown in Fig.~\ref{fig3}. The system shows three distinct phases: a gas phase, which is characterized by a vanishing $n^{\rm eq.}_{\rm 2D}$ and absence of order, an unmodulated liquid phase, which has a finite equilibrium 2D density at which $E/N < \hbar \omega/2$ with absent order and finally a liquid, modulated phase, where $E/N < \hbar \omega/2$ for $n^{\rm eq.}_{\rm 2D}$ and the system shows modulations. We observe there is a maximum value of the scattering length for the emergence of modulations at the equilibrium density, which lies around $a/a_{\rm dd} \simeq 0.76$ for our parameters of choice, or $a \simeq 99.4 a_0$ for $^{164}$Dy atoms. This phenomenology is consistent to the one observed in the 2D case~\cite{staudinger23:PRA}. However, the comparison between the results of Fig.~\ref{fig3} with the phase diagram provided in Ref.~\cite{staudinger23:PRA} shows relevant qualitative differences drawn from the finite trapping strength along the $z$-axis. While in the strict 2D case the system can transition directly from the gas phase to the modulated liquid, we find that the phase boundary of the gas phase is always connected to the unmodulated liquid for our parameters of choice. The difference stems from the dimensional nature of the modulated phase, since in all cases it has a 3D peak density significantly larger than the threshold density $n_{\rm lim} r_0^3 \simeq 0.06$ at which excitations along the $z$ axis get significantly populated (see Sec.~\ref{sec:theory}). This means that this state is three-dimensional.

\begin{figure}[t]
\centering
\includegraphics[width=0.8\linewidth]{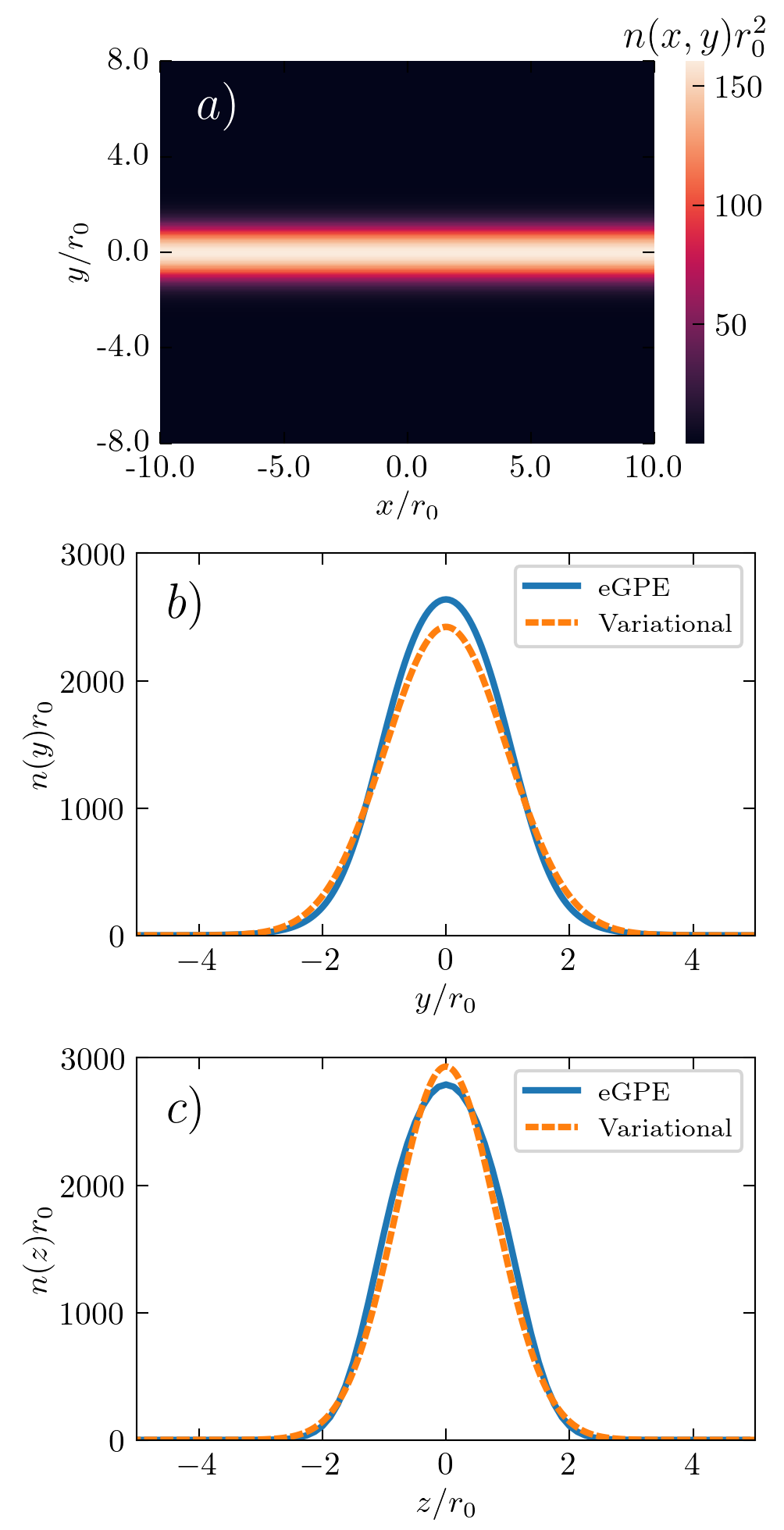}
\caption{$a)$ 2D column density $n(x,y) = \int dz \abs{\psi({\bf r})}^2$ of a single stripe of the eGPE solution. $b)$,$c)$ Column densities $n(y) = \int dx dz \abs{\psi({\bf r})}^2$ ($b)$) and $n(z) = \int dx dy \abs{\psi({\bf r})}^2$ ($c)$) from the ground state solution of the eGPE (Eq.~\ref{eGPE}) and the gaussian ansatz at the variational energy minimum (Eq.~\ref{gaussian}). The scattering length and tilting angle are $a/a_{\rm dd} = 0.7$ and $\alpha = 75.92^o$, respectively, while the 2D density is set to $n_{\rm 2D} r_0^2 = 20$. }
\label{fig5}
\end{figure}

Another major qualitative difference between our phase diagram and that of Ref.~\cite{staudinger23:PRA} is the fact that the liquid phases in the fully 2D case appear only when the system hosts a two-body bound state. In contrast, we find the liquid states with a model potential that does not host any two-body bound state, meaning that the self-bound nature of the system arises due to many-body effects. This has important consequences in the formation of self-bound droplet states in the finite size system. While in the 2D case of Ref.~\cite{staudinger23:PRA} any many-body state in the liquid phase supports the formation of a self-bound droplet, in our case there is a varying critical number of particles for droplet formation. We study the formation of dipolar clusters in the finite system in Sec.~\ref{sec:finite_size}. It must be remarked, however, that the model interaction potential employed in Ref.~\cite{staudinger23:PRA} differs from ours in the short range part, since a hard core potential of the form $C/r^{12}$ (with $r$ the inter-particle distance and $C$ a constant) is used in the reference instead of the contact short range potential of Eq.~\ref{pseudopot}. However, for all the calculations considered in this work, the system lies in the dilute regime, since $n_{\rm peak} a^3 < 10^{-3}$. Hence, its physics are independent on the specific short range details of the interaction.

\begin{figure}[t]
\centering
\includegraphics[width=0.8\linewidth]{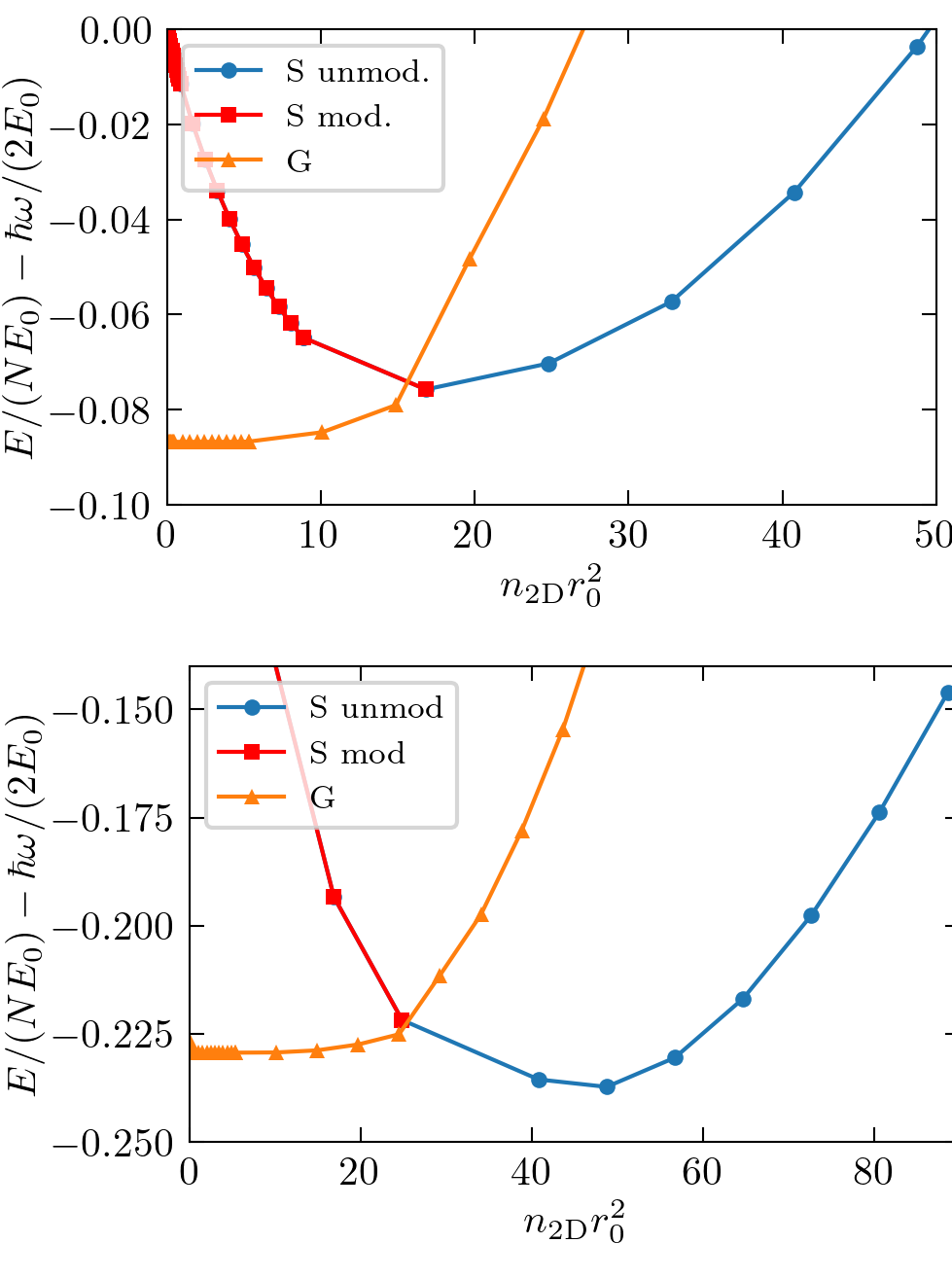}
\caption{Energy per particle for the senoidal (see Eq.~\ref{senoidal}, labeled by ``S'' in the figure) and gaussian (see Eq.~\ref{gaussian}, labelled by ``G'' in the figure) ansatz as a function of the 2D density for $\alpha=77.35^o$ (top, striped liquid region) and $\alpha=81.36^o$ (bottom, unmodulated liquid region). The labels ``mod'' and ``unmod'' denote a modulated (unmodulated) senoidal ansatz solution. The calculations correspond to the infinite system in the $x$-$y$ plane. The value of the scattering length is set to $a/a_{\rm dd} = 0.7$. The red squares indicate the values at which the senoidal ansatz shows a density modulation in the variational energy minimum. We have subtracted the energy of the non-interacting system, $E_{\rm NI}/(N E_0) = \hbar \omega/(2 E_0)$.}
\label{fig6}
\end{figure}

We have analyzed the transitions between the different regions of the phase diagram of Fig.~\ref{fig3} for two values of the scattering length, $a/a_{\rm dd} = 0.6$ and $0.7$. To do so, we have looked at the equilibrium 2D density $n^{\rm eq.}_{\rm 2D}$ in the gas-liquid transition and at the contrast of the 3D density, $C = \frac{n_{\rm max.} - n_{\rm min.}}{n_{\rm max.} + n_{\rm min.}}$ in the transition between the modulated and unmodulated liquid regions. Results are shown in Fig.~\ref{fig4}. For both values of the scattering length, the equilibrium density 2D density changes continuously from $n^{\rm eq.}_{\rm 2D} = 0$ in the gas phase to $n^{\rm eq.}_{\rm 2D}\neq 0$ in the liquid phase, meaning that the liquid state emerges continuously. For this reason, we only report the case with $a/a_{\rm dd} = 0.7$. In regards to the transition in and out of the striped region, the contrast can change either abruptly ($a/a_{\rm dd} = 0.6$) or continuously ($a/a_{\rm dd} = 0.7$) when entering the striped region by increasing the tilting angle, as shown in the figure. This is because, for $a/a_{\rm dd} = 0.6$, the gaussian ansatz becomes energetically favorable at the equilibrium density over an unmodulated state described by the senoidal ansatz. In contrast, for $a/a_{\rm dd} = 0.7$, the senoidal ansatz, for which modulations arise continuously, is preferable to the gaussian one over a larger range of densities. Exiting the modulated region by increasing the tilting at constant scattering length causes to contrast to abruptly vanish for both cases, and hence it is not shown in the figure.

\begin{figure}[t]
\centering
\includegraphics[width=0.8\linewidth]{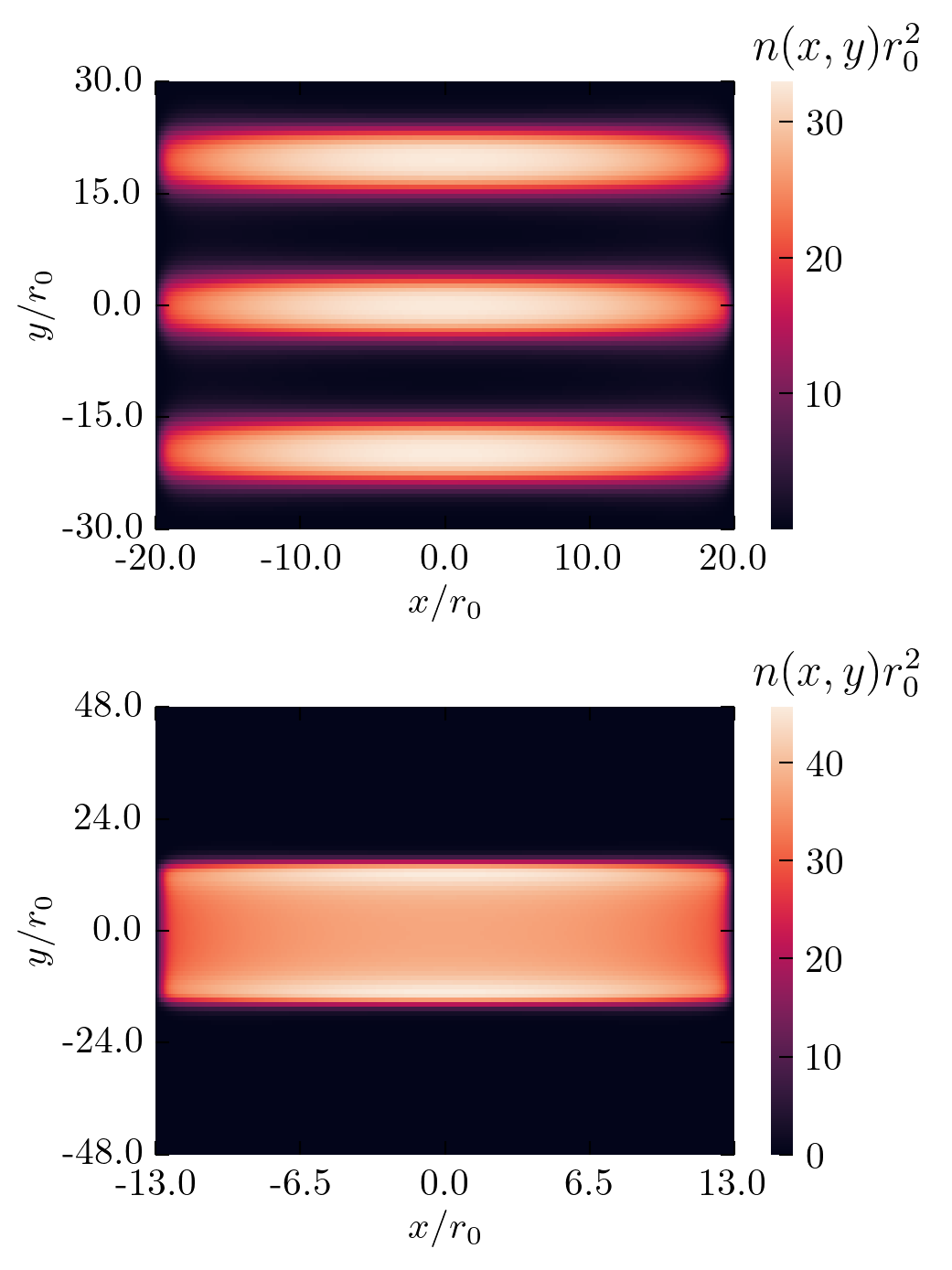}
\caption{Integrated density  $n(x,y) = \int dz \abs{\psi({\bf r})}^2$ for the ground state of the flattened dipolar system in a box trap. The tilting angle and the lengths of the box trap are $\alpha=77.35^o$, $L_x/r_0 = 41.7$, $L_y/r_0 = 61.5$ (top) and $\alpha=81.36^o$, $L_x/r_0 = 26.3$, $L_y/r_0 = 96.3$ (bottom) while, in both cases, $a/a_{\rm dd} = 0.7$, $N = 30000$.}
\label{fig7}
\end{figure}

\begin{figure}[t]
\centering
\includegraphics[width=\linewidth]{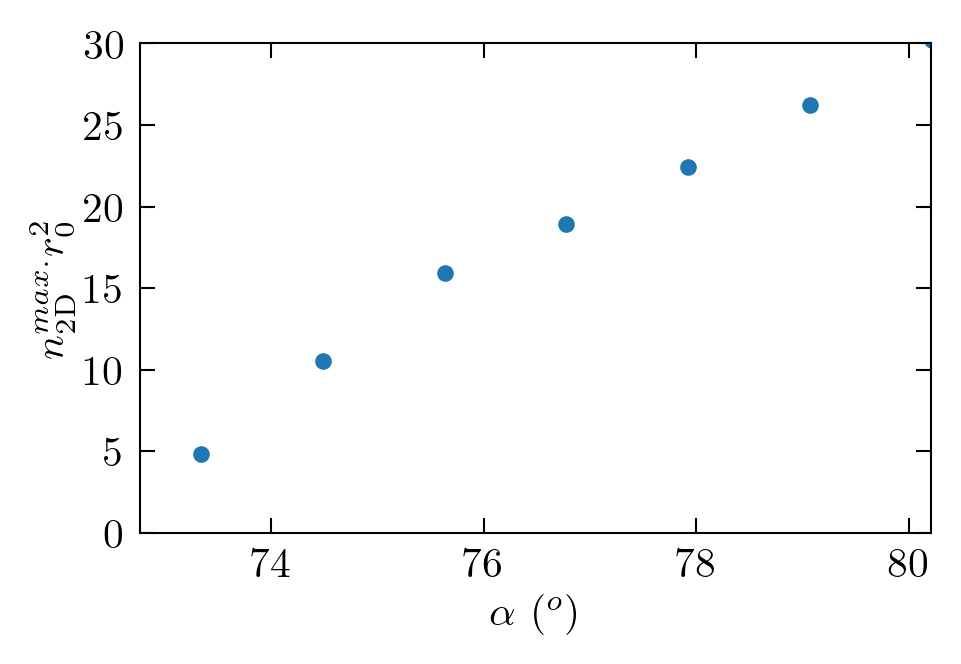}
\caption{Maximum 2D density at which a striped state appears as the variational energy minimum in the infinite system vs tilting angle. We set $a/a_{\rm dd} = 0.7$. The values of the tilting are chosen within the striped liquid region of Fig.~\ref{fig2}.}
\label{fig8}
\end{figure}

In order to evaluate the effect of increasing the trapping strength in our results, we perform a variational calculation for $a/a_{\rm dd}=0.7$ using a trapping strength of $\omega' = 2\pi \times 90600$ Hz, with an associated harmonic oscillator length of $l = 26$ nm, around the order of magnitude of that employed in the experiment of Ref.~\cite{li_du2024:science}. Such an increase of the harmonic oscillator frequency has an important effect on
$n_{\rm lim}$, which increases by almost two orders of magnitude ($n_{\rm lim} r_0^2 \simeq 3$). The results indicate that for the trapping strength $\omega'$, the system only hosts the gas and unmodulated liquid phases at the equilibrium density, with the transition happening at $\alpha = 73.63^o$. This corresponds to the behaviour of the phase diagram at larger values of $a/a_{\rm dd}$ for $\omega$, which implies that, effectively, the modulated liquid phase is pushed to lower values of the scattering length.

\begin{figure*}[t]
\centering
\includegraphics[width=\linewidth]{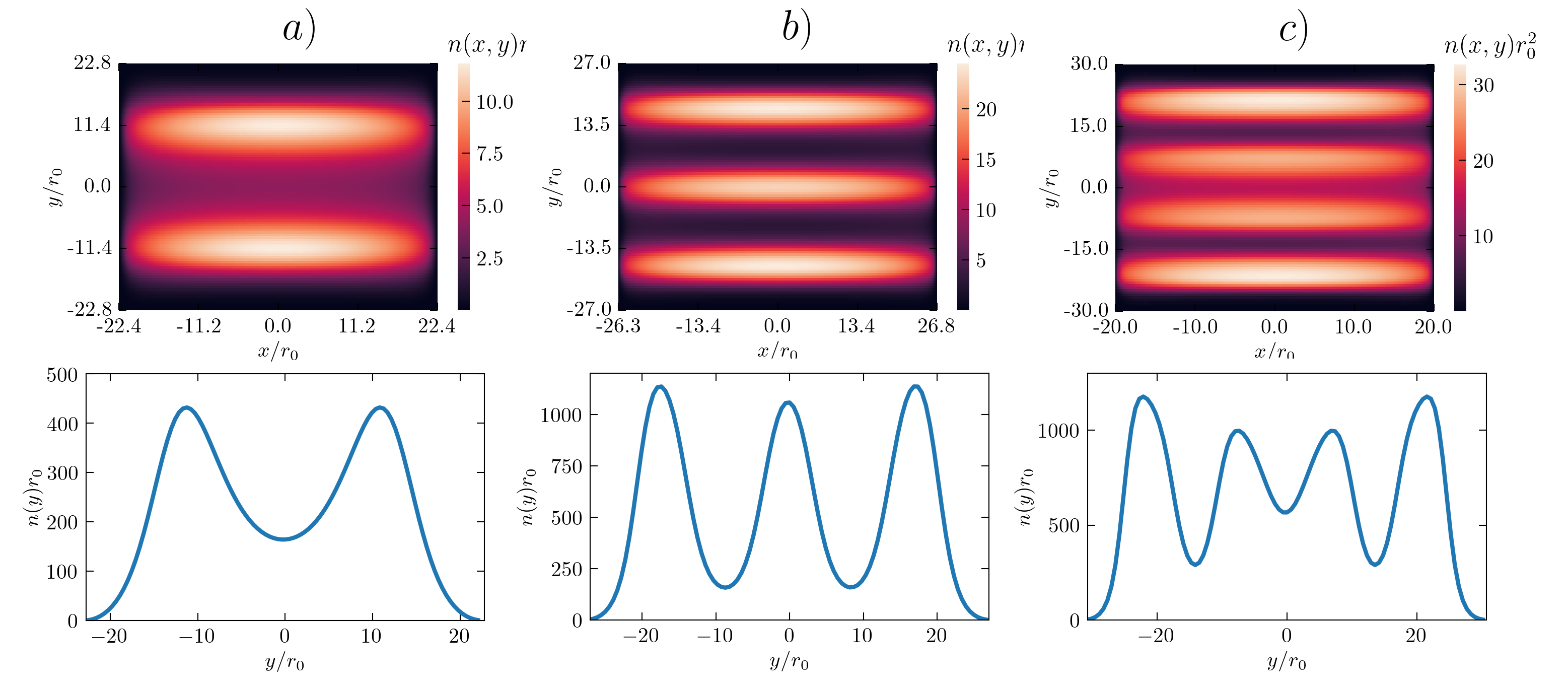}
\caption{Potentially supersolid states: integrated densities  $n(x,y) = \int dz \abs{\psi({\bf r})}^2$ (top) and column densities $n(y) = \int dx dz \abs{\psi({\bf r})}^2$ (bottom) for the ground state of the flattened dipolar system in a box trap. The parameters are $N=10000$, $\alpha=74.48^o$, $L_x/r_0 = 45.6$, $L_y/r_0 = 45.6$ (a)), $N=28000$, $\alpha=75.63^o$, $L_x/r_0 = 54.3$, $L_y/r_0 = 54.3$ (b)) and $N=40000$, $\alpha=77.35^o$, $L_x/r_0 = 41.1$, $L_y/r_0 = 61.5$ (c)). In all cases $a/a_{\rm dd} = 0.7$.}
\label{fig9}
\end{figure*}

\begin{table*}
    \begin{center}
        \begin{tabular}{ | c | c | c | c | }
        \hline
            ($N$, $\alpha$) & $f_s^{+}$ & $f_{s}^{-}$ & $\tilde{L}_y/r_0$ \\ \hline
            ($10000$, $74.48^o$) & $0.861$ & $0.860$ & $32.8$ \\ \hline
            ($28000$, $75.63^o$) & $0.580$ & $0.577$ & $46.8$ \\ \hline
            ($40000$, $77.35^o$) & $0.677$ & $0.673$ & $53.2$ \\ \hline
        \end{tabular}
    \end{center}
    \caption{Upper ($f_s^+$) and lower ($f_s^-$) Leggett's bounds for the superfluid fraction (see Eqs.~\ref{legget_upper} and~\ref{legget_lower}) of the ground states shown in Fig.~\ref{fig9}, together with the size of the integration domain along the $y$ axis in the computation of the bounds, which is given by $\tilde{L}_y$.}
    \label{table1}
\end{table*}

\begin{figure}[t]
\centering
\includegraphics[width=\linewidth]{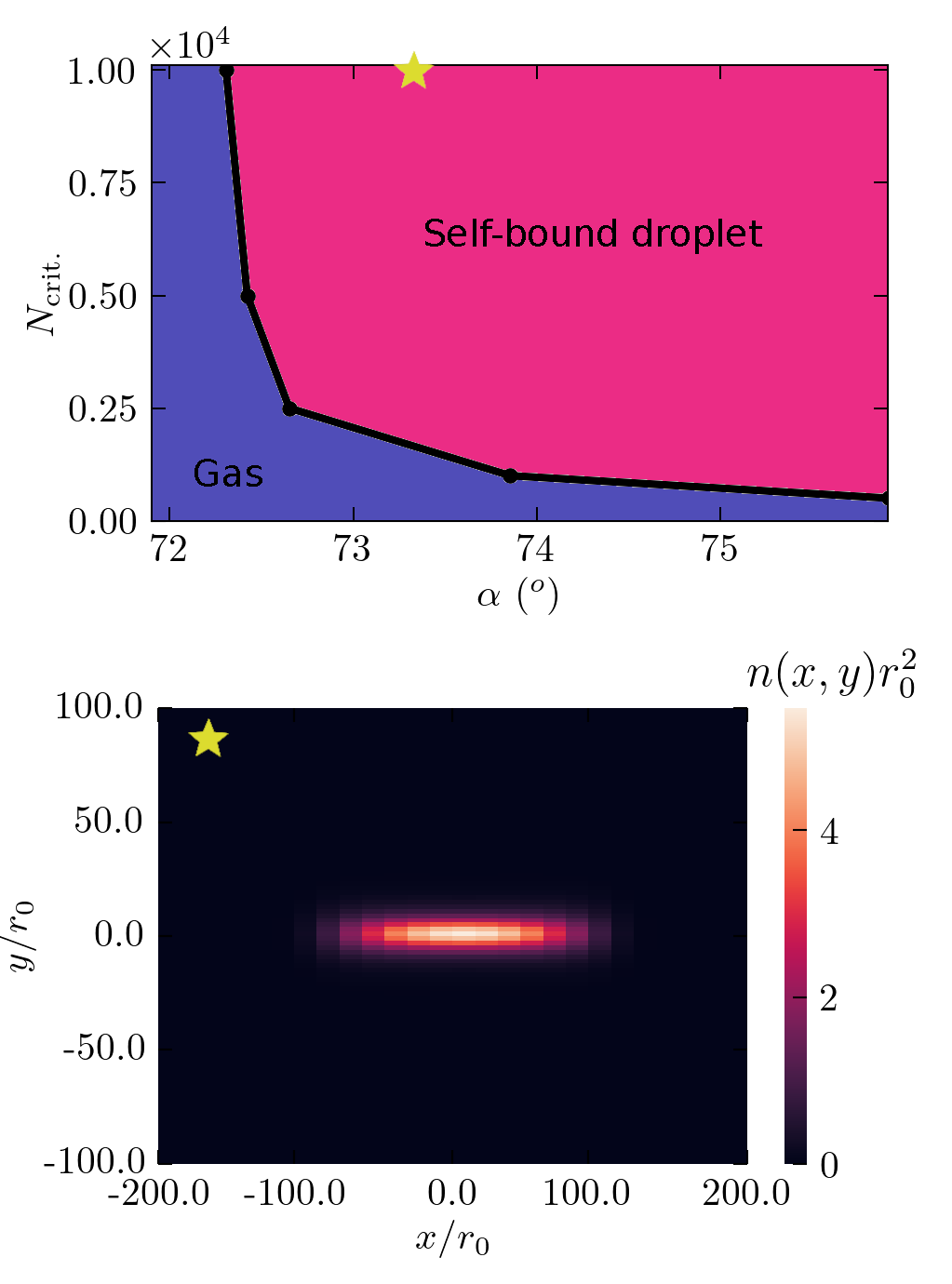}
\caption{Top: critical number for stable droplet formation vs tilting angle for $a/a_{\rm dd} = 0.7$. In all cases where a droplet is formed, the length scales of the harmonic trap in the $x$-$y$ plane are significantly larger than the size of the droplet. Bottom: integrated density $n(x,y) = \int dz \abs{\psi({\bf r})}^2$ for a self-bound droplet of $N=10000$ atoms at $\alpha = 73.68^o$.}
\label{fig10}
\end{figure}

In order to benchmark our variational calculations, we run simulations to solve the eGPE of Eq.~\ref{eGPE} and compare the outcome with the variational results in two cases. Firstly, we set $a/a_{\rm dd} = 0.6$, $\alpha=75.92^o$, $n_{\rm 2D} r_0^2 = 20$ and compute the ground state wave function and energy of the system. The eGPE yields a striped state with $E/N - \hbar \omega/2 = -0.47662 E_0$ and a stripe spacing of $L_y/r_0 = 21$ while the variational calculation yields a striped gaussian state with $E/N - \hbar \omega/2 = -0.44373 E_0$ with $L_y/r_0=18.5$, providing reasonable agreement. We show a comparison of the column densities $n(y) = \int dx dz \abs{\psi({\bf r})}^2$ and $n(z) = \int dx dy \abs{\psi({\bf r})}^2$ obtained with both methods in Fig.~\ref{fig5}, together with the 2D column density $n(x,y) = \int dz \abs{\psi({\bf r})}^2$ of a stripe obtained from the eGPE. Secondly, we perform eGPE simulations for $a/a_{\rm dd} = 0.6$, $\alpha=67.9^o$, $69.9^o$ at $n_{\rm 2D} r_0^2 = 0.56$ and find results compatible with the variational calculations of Fig.~\ref{fig3}, mainly an unmodulated, gas state at $\alpha=67.9^o$ and a modulated, liquid state at $\alpha=69.9^o$. The number of points of the spatial grid employed in the numerical simulations are $\{N_x,N_y,N_z\} = \{80, 400, 80\}$ with an imaginary time step of $\Delta \tau = 0.00025 \hbar/E_0$ and $N_{\rm it} = 200000$ iterations.

\subsection{\label{sec:finite_size}Finite system}

We now turn our attention to the finite system, where we consider a trapping potential of the form $U({\bf r}) = \frac{1}{2} m \omega^2 z^2 + U_{\rm box}(x,y)$. As in the infinite case, we set the trapping strength along the $z$-axis to $\omega = 2 \pi \times 906$ Hz in all calculations. The box trap potential is characterized by parameters $L_x$ and $L_y$ and is given by
\begin{align}
\small
 U_{\rm box}(x,y) =
 \begin{cases}
  0 &\text{ if } x \in [-L_x/2, L_x/2],y \in [-L_y/2, L_y/2]  \\
  \infty &\text{ otherwise }
 \end{cases}
 \label{box_trap}
\end{align}

In order to obtain the ground state of the system, we solve the eGPE given in Eq.~\ref{eGPE} by simulating imaginary time evolution. All the states computed for the finite system fulfill $n_{\rm peak} \gg n_{\rm lim}$, meaning that we employ the beyond mean-field energy functional of Eq.~\ref{fluc_2}. We first focus on the realization of the striped liquid phase in experimental conditions. In this sense, the results of Fig.~\ref{fig3} are particularly useful. This is illustrated in Fig.~\ref{fig6}, where we show the behaviour of $E/N$ for the infinite system for two points in the liquid phase: $\{a, \alpha\} = \{ 0.7, 77.35^o \}$ (modulated liquid) and $\{a, \alpha\} = \{ 0.7, 81.36^o \}$ (unmodulated liquid). Notice that in the unmodulated liquid region, the system's energetically preferable state is an unmodulated liquid for $n_{\rm 2D} \geq n^{\rm eq.}_{\rm 2D}$. This means that, when experimentally realizing the finite system under the box trapping, the system is unmodulated for $n_{\rm 2D} \geq n^{\rm eq.}_{\rm 2D}$ while it converges to an unmodulated state also for $n_{\rm 2D} < n^{\rm eq.}_{\rm 2D}$, since particles minimize their energy by bunching and producing a state with 2D density as close to $n^{\rm eq.}_{\rm 2D}$ as possible. On the other hand, this does not happen in the modulated liquid region. To illustrate this phenomenology, we perform simulations at the two aforementioned points, $\{a, \alpha\} = \{ 0.7, 77.35^o \}$ and $\{ 0.7, 81.36^o \}$ setting $N = 30000$ and $\{ L_x/r_0, L_y/r_0\} = \{ 41.7, 61.5 \}$ in the former case and $\{ L_x/r_0, L_y/r_0\} = \{ 26.3, 96.3 \}$ in the latter. The lengths of the box trap are chosen to fix a 2D density where the variational calculation in the infinite case leads to density modulations, with $L_y$ chosen to be commensurate with their spatial period. We show the results in Fig.~\ref{fig7}. As can be seen from the results, we see the emergence of a striped state for $\{a, \alpha\} = \{ 0.7, 77.35^o \}$ but not so for $\{ 0.7, 81.36^o \}$, where the atoms bunch into an unmodulated state with higher 2D density to minimize their energy. Remarkably, in this case, the atoms do not occupy the full volume of the trap. In view of these results, it is thus important to characterize the maximum 2D density at which stripes emerge in the infinite system within the modulated liquid region, since setting an experiment with higher 2D density would destroy modulations. Thus, we report this quantity in Fig.~\ref{fig8} for $a/a_{\rm dd} = 0.7$.

Next, we move on to the discussion on supersolidity. In qualitative terms, a modulated state should feature a cloud of atoms surrounding the density modulations in order for global phase coherence to exist, and thus for supersolidity to emerge. We show in Fig.~\ref{fig9} the integrated density $n(x,y)$ of three examples of supersolid states, which involve parameters currently achievable in experiments. In order to quantify its superfluid fraction, we employ Leggett's upper and lower bounds for the superfluid fraction, which are given by~\cite{leggett70:prl,leggett98:jsph}
\begin{align}
 f_s^+ &= \frac{\tilde{L}^2_y}{\tilde{N}} \left[ \int_{-\tilde{L}_y/2}^{\tilde{L}_y/2} \frac{dy}{n(y)} \right]^{-1} \label{legget_upper} \\
 f_s^- &= \frac{\tilde{L}^2_y}{\tilde{N}} \int_{-\infty}^{\infty} \int_{-L_x/2}^{L_x/2} dz dx \left[ \int_{-\tilde{L}_y/2}^{\tilde{L}_y/2} \frac{dy}{\abs{\psi({\bf r})}^2} \right]^{-1} \label{legget_lower} \ ,
\end{align}
where $f_s^+$ and $f_s^-$ are Leggett's upper and lower bounds, respectively, and
$n(y) = \int dx dz \abs{\psi({\bf r})}^2$. It is important to remark that, in Eqs.~\ref{legget_upper} and~\ref{legget_lower}, the quantity $\tilde{L}_y$ does not correspond to the simulation box size along the $y$ axis, which is denoted by $L_y$. In much the same way, $\tilde{N}$ is given by $\tilde{N} = \int_{-\tilde{L}_y/2}^{\tilde{L}_y/2} dy n(y)$. This is because the Leggett's bounds are formulated for an infinite system, and therefore, their estimation of the superfluid fraction corresponds to an infinite system formed by periodic repetitions of the density distribution within the domain $[-\tilde{L}_y/2, \tilde{L}_y/2]$. Thus, to rightfully estimate the superfluid fraction of our finite size states, we need the pattern of the column densities $n(y)$ shown in Fig.~\ref{fig9} to repeat itself consistently with the modulations present away from the edges. Because of this, $\tilde{L}_y$ is chosen such that $y=-\tilde{L}_y/2$ is the minimum value of $y$ for which $n(-\tilde{L}_y/2) = n(\tilde{L}_y/2) = \text{Min.}\{ n(y) \}$, with $\text{Min.}\{ n(y) \}$ the \textit{local} minimum closest to the edges of the trap. We show in Table~\ref{table1} the results for the Leggett's upper and lower bounds for the three states shown in Fig.~\ref{fig9} together with the chosen values of $\tilde{L}_y$. The results indeed confirm that the three states shown are supersolid. Moreover, due to both bounds being remarkably close, we are able to provide a quantitative estimate of the superfluid fraction at the mean field level. The closeness between both bounds in the shown cases is not unexpected, since the bounds are equivalent for a density distribution that factorizes as $n({\bf r}) = n_y(y) n_{\perp} (x,z)$~\cite{perez25:pra}, as it is the case for the density distribution shown in Fig.~\ref{fig9} away from the box edges.

One concern for the realization of these supersolid states states is the peak density, as a high value can induce excessive three-body losses. In all the states reported in Fig.~\ref{fig9}, the peak 3D density is of the order $n_{\rm peak} r_0^{3} \sim \order{10}$, which means $n_{\rm peak} \sim 5 \times 10^{14}$ cm$^{-3}$. This value lays within experimental reach~\cite{Kadau:2016cb}, such that three-body losses do not translate into an excessively short life-time~\cite{zhang21}. These results may thus guide potential future experiments in the realization of supersolid stripes, which were not observed in the experiment of Ref.~\cite{wenzel_2017:PRA}. The results of Figs.~\ref{fig7} and~\ref{fig9} have been obtained for a spatial grid with $\{N_x,N_y,N_z\} = \{300, 100, 40\}$ points with an imaginary time step of $\Delta \tau = 0.0008 \hbar/E_0$ and $N_{\rm it} = 200000$ iterations.

Finally, we address the formation of dipolar droplets. The appearance of a liquid phase in the thermodynamic limit (see Fig.~\ref{fig3}) is indicative of the emergence of self-bound liquid droplets in the finite size system. In the absence of a trap in the $x$-$y$ plane, dipoles minimize their energy by forming a droplet elongated along the $x$-axis (which can be understood as a single, finite stripe)~\cite{wenzel_2017:PRA}. However, and distinctly to what happens in the strict 2D regime, there exist a minimum number of particles greater than two necessary for the emergence of a many-body bound state. We refer to this quantity as the critical number $N_{\rm crit.}$. We show in Fig.~\ref{fig10} $N_{\rm crit.}$ as a function of the tilting angle $\alpha$ for $a/a_{\rm dd}=0.7$, as well as a sample of the column density $n(x,y)$. We have replaced the box trap defined in Eq.~\ref{box_trap} by a weak harmonic trap in the $x$-$y$ plane $U(x,y) = \frac{1}{2} m \left( \omega_x^2 x^2 + \omega_y^2 y^2 \right)$, which has no effect on our results. The harmonic frequencies are $\hbar \omega_x/E_0 =2.20 \times 10^{-6}$ and $\hbar \omega_y/E_0 =8.83 \times 10^{-6}$ for $N > 1750$ and $\hbar \omega_x/E_0 = 3.91 \times 10^{-6}$ and $\hbar \omega_y/E_0 = 3.53 \times 10^{-5}$ for $N \leq 1750$. These trapping frequencies have been chosen such that the characteristic length scale of the trap is significantly larger than the size of the self-bound droplets, such that the contribution to the energy by the trapping term is neglectable. It must be remarked that we are only considering the formation of stable droplets, for which $E/N < \hbar \omega/2$, although metastable droplets for which $E/N > \hbar \omega/2$ may arise for slightly lower tilting angles/particle numbers. On a technical note, the swap from a box trap to a harmonic trap responds to the need to compute the kinetic energy term in this case via a Fast-Fourier Transform (FFT) algorithm instead of doing so by using a finite differences formula. This is because this alternative offers a better convergence with the number of spatial grid points. While in the rest of the calculations of our work we use a number of grid points for which the two methodologies give equivalent results, fine numerical accuracy is specially important when determining the formation of stable droplets with the aforementioned energy criterion. The FFT kinetic energy computation could have also been applied for the box trap, but our particular implementation of this trapping potential relies on setting to zero the points of the wave function at the ends of the spatial grid in the kinetic energy calculation through finite differences. The results of Fig.~\ref{fig10} have been obtained for a spatial grid with $\{N_x,N_y,N_z\} = \{50, 100, 40\}$ points with an imaginary time step of $\Delta \tau = 0.0005 \hbar/E_0$ and $N_{\rm it} = 300000$ iterations.

The solution of the eGPE yields a minimum tilting angle for stable droplet formation of $\alpha = 72.31^o$ for the parameters considered, which lays close to the tilting for the liquid-gas phase transition for the infinite system (see Fig.~\ref{fig3}), $\alpha = 71.68^o$. The slight discrepancy between both values most likely arises from the variational approximation employed in the calculations of the infinite system. The realization of these self-bound droplets in experiments with flattened geometries is also feasible, since the peak density fulfills $n_{\rm peak}< 10^{15}$ cm$^{-3}$ for $\alpha<77.92^o$, meaning that three-body losses should be manageable in the regime $\alpha \in (72.31^o, 77.92^o)$.

\section{\label{sec:conclusions}Conclusions}

In conclusion, we have computed the phase diagram at the 2D equilibrium density of a system of tilted dipolar bosons under a realistic harmonic confinement along the $z$-axis. We have done so by performing a variational calculation, which we have benchmarked with the solution of the extended Gross-Pitaevskii equation. We have included the effect of beyond mean-field corrections accounting for the discretization of excitation energies along the trapping direction. The diagram hosts three distinct phases in the thermodynamic limit: a gas phase, an unmodulated liquid phase and a striped, modulated liquid phase. Remarkably, and distinctly to the phenomenology in the strict 2D limit, the liquid phases emerge even if the two-body interaction does not host a two-body bound state. We have characterized the transitions between these phases, and have sketched the effect of increasing the trapping strength. We have also performed simulations for the finite system, and have connected the phenomenology in the thermodynamic limit with the results. We have illustrated several cases within experimental reach where the realization of potentially supersolid striped states should be possible, and have also characterized the formation of self-bound, stable liquid droplets for a realistic value of the scattering length of $a = 91.56 a_0$ for $^{164}$Dy atoms, and for realistic particle numbers.

Our results should serve to motivate and guide potential future experiments in the study of dipolar systems in flattened geometries, in the quasi-2D limit, especially in regards to the realization of liquid phases and supersolid striped states. The proper quantification of the superfluid properties of such states is a matter for future work, where exact quantum Monte Carlo methods could be applied to exactly obtain the superfluid fraction across stripes. Moreover, the role played by dimensionality in thermal effects should also be more extensively characterized in order to check if the surprising and counter-intuitive promotion of supersolidity by heating seen in three dimensions~\cite{baena22,baena24,He2024} is robust to dimensionality.

We acknowledge financial support from Ministerio de Ciencia e 
Innovaci\'on
MCIN/AEI/10.13039/501100011033
(Spain) under Grant No. PID2023-147469NB-C21  and
from AGAUR-Generalitat de Catalunya Grant No. 2021-SGR-01411.

\appendix

\section{\label{sec:dens_lim}Determination of $n_{\rm lim}$ and the fitting constants of Eqs.~\ref{fluc_1} and~\ref{fluc_2}}

In order to determine the limiting 3D density $n_{\rm lim}$ from Eqs.~\ref{fluc_1} and~\ref{fluc_2} in the main text, we numerically compute the beyond mean-field correction to the chemical potential, $ \mu_{\rm BMF}$, which can be obtained from the solution of the Bogoliubov de-Gennes equations in the usual manner. The only remarks are the use of the potential in Fourier space employed in Ref.~\cite{Zin_2021} and the fact that the excitations along the $z$-axis are discrete. Since this quantity is ultraviolet divergent, as it is well known, we employ the standard renormalization procedure used in three-dimensions, analogously to the renormalization of the energy density performed in Ref.~\cite{Zin_2021}. We then choose $n_{\rm lim}$ such that the ratio $ \mu_{\rm BMF} / \mu_{\rm 3D} < 1.3$ for $n>n_{\rm lim}$ for all the calculations considered in this work. Here, $\mu_{\rm 3D}$ is the usual beyond mean-field correction for a dipolar system in three-dimensions, i.e.
\begin{align}
 \mu_{\rm 3D} =\frac{32 g \sqrt{a^3}}{3 \sqrt{\pi}} \mathcal{Q}_{5}\left( \frac{a_{\rm dd}}{a} \right) n^{3/2},
\end{align}
where $g = 4 \pi \hbar^2 a/m$ and and
$Q_5(\varepsilon_{\mathrm{dd}}) = \dfrac{1}{2} \displaystyle \int_{0}^{\pi}
\mathrm{d} \alpha \sin\alpha \left[1 + \varepsilon_{\rm dd} (3 \cos^2 \alpha -
1)\right]^{5/2}$.. The constants of Eqs.~\ref{fluc_1} and~\ref{fluc_2} are then obtained by fitting the correction of the chemical potential to
\begin{align}
  \mu^{\rm low}_{\rm BMF}(n) &= \tilde{C}_1 n + \tilde{C}_2 n^2 + \tilde{C}_3 n^{3/2} & n < n_{\rm lim} \label{fluc_1_mu} \\
  \mu^{\rm high}_{\rm BMF}(n) &= \tilde{C}_4 n^{3/2} & n > n_{\rm lim} \label{fluc_2_mu} \ .
\end{align}

\begin{figure}[t]
\centering
\includegraphics[width=\linewidth]{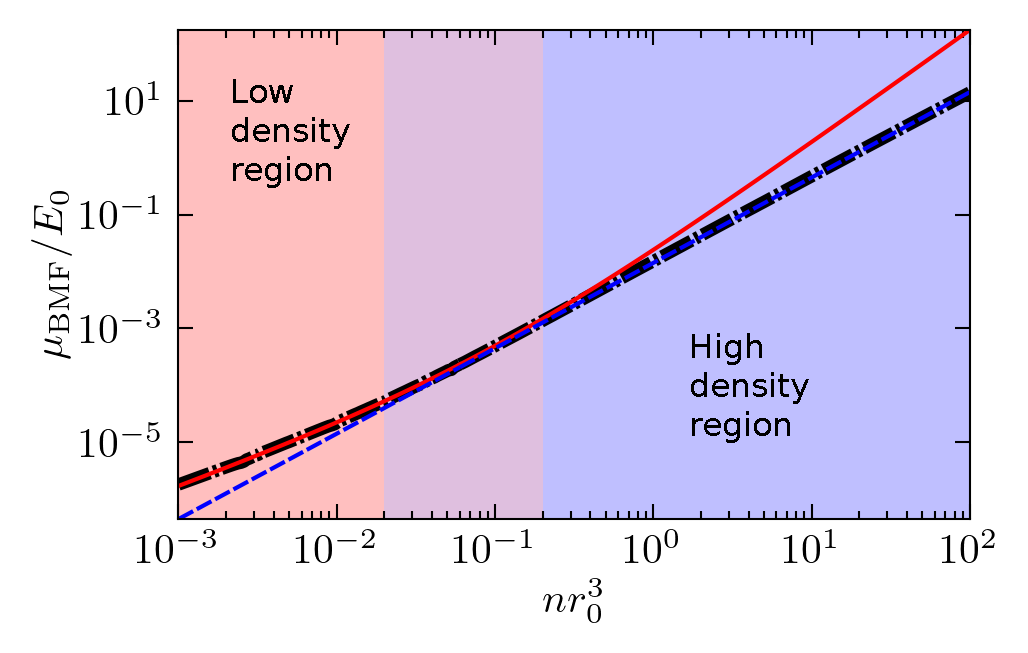}
\caption{Beyond mean-field correction to the chemical potential obtained from the solution of the Bogoliubov de-Gennes equations (black dash-dotted line) as a function of the 3D density. The solid and dashed lines correspond to the fits of Eqs.~\ref{fluc_1_mu} and~\ref{fluc_2_mu}, respectively. The labels ``Low density region'' and ``High density region'' make reference to the superscripts ``low'' and ``high'' of Eqs.~\ref{fluc_1_mu} and~\ref{fluc_2_mu}. The parameters are $a/a_{\rm dd} = 0.7$ and $\alpha=75.92^o$.}
\label{fig_app}
\end{figure}

Then, using the relation $\eval{ \pdv{\epsilon_{\rm BMF}}{n} }_V = \mu_{\rm BMF}$ we obtain $C_1 = \tilde{C}_1/2$, $C_2 = \tilde{C}_2/3$ and $C_{3,4} = 2\tilde{C}_{3,4}/5$. We show in Fig.~\ref{fig_app}, the ratio $ \mu_{\rm BMF} / \mu_{\rm 3D}$ as a function of the density for $a/a_{\rm dd} = 0.7$, $\alpha=75.91^o$, $\hbar \omega / E_0 = 1$, together with the fits of Eqs.~\ref{fluc_1_mu} and~\ref{fluc_2_mu}. For these parameters, we find $n_{\rm lim} r_0^3 = 0.06$.

\begin{table*}
    \begin{center}
        \begin{tabular}{ | c | c | c | c | }
        \hline
            State & $n_{\rm 2D} r_0^2$ & $E/(N E_0) - \hbar \omega/(2 E_0)$ & $L_y/r_0$ \\ \hline
            Triangular & $1$ & $-0.456$ & $600$ \\ \hline
            Triangular & $0.5$ & $-0.472$ & $1040$ \\ \hline
            Triangular & $0.25$ & $-0.483$ & $2000$ \\ \hline
            Triangular & $0.125$ & $-0.488$ & $4000$ \\ \hline
            Stripe & $0.5$ & $-0.490$ & $880$ \\ \hline
        \end{tabular}
    \end{center}
    \caption{Energies per particle of the converged triangular states and the optimal stripe state for $a/a_{\rm dd} = 0.6$, $\alpha = 75.9^o$. The triangular states are obtained by using the ansatz of Eq.~\ref{triangular} as the initial state of the eGPE simulations. We also report the box size length $L_y/r_0$ that yields the optimal energy per particle.}
    \label{table2}
\end{table*}

\section{\label{sec:triangular}Existence of triangular solid and honeycomb ground states}

Previous works which consider lower values of the tilting angle show the emergence of both a triangular and honeycomb states as the ground state of the system for a given range of densities~\cite{zhang19,zhang21,cinti2025}. In the variational calculations performed in this work, we have not consider neither of these states. However, we can argue that they do not correspond to the ground state of the system at the equilibrium density for the parameters of choice of our work.

Firstly, in regards to the honeycomb state, increasing the tilting angle shrinks the domain of its region in the ($a/a_{\rm dd}, n_{2D}$) phase diagram of the system, pushing it to densities considerably higher than the ones obtained in this work~\cite{cinti2025}. Because of this, it is safe to ignore it for the high tilting angles considered here ($\alpha > 69^{o}$). Moreover, this assumption is consistent with the results of Ref.~\cite{staudinger23:PRA}, where $\alpha \in [57^o, 69^o]$ and the honeycomb state is not found in the phase diagram of the system computed at the equilibrium density.

In regards to the triangular solid phase, the results of Ref.~\cite{cinti2025} show that the phase domain of this state is pushed to lower values of the scattering length and lower densities as the tilting increases. Thus, if it is not the ground state at the equilibrium density for the lowest value of the scattering length considered ($a/a_{\rm dd} = 0.6$), we can rule out its existence for the purposes of this work. To check this, we have numerically solved the eGPE (see Eq.~\ref{eGPE}) for $a/a_{\rm dd} = 0.6$ and different values of the density and for tilting angles $\alpha = 71.9^o, 75.9^o, 77.9^o$. To guarantee the convergence into a triangular solid state, we start the simulation from an initial state given by
\begin{align}
 &\psi_{0}(\bf{r}) = \mathcal{N} \left( 1 + 0.3 \cos({\bf k_1} {\bf r_{\perp}}) \nonumber \right. \\
 & \left. + 0.3 \cos({\bf k_2} {\bf r_{\perp}}) + 0.3 \cos({\bf k_3} {\bf r_{\perp}}) \right)^{1/2} \nonumber \\
 &\times e^{-\frac{1}{2} \omega^2 z^2}
 \label{triangular}
\end{align}
which corresponds to a triangular lattice of clusters. Here, $\mathcal{N}$ is a normalization constant such that $N = \int d{\bf r} \abs{\psi_{0}(\bf{r})}^2$. The momentum vectors are ${\bf k_1} = k_0 \hat{x}$, ${\bf k_2} = -k_0 \left( \cos(\pi/3) \hat{x} + \sin(\pi/3) \hat{y} \right)$ and ${\bf k_3} = k_0 \left( -\cos(\pi/3) \hat{x} + \sin(\pi/3) \hat{y} \right)$, with $\hat{x}$ and $\hat{y}$ the unitary vectors along the $x$ and $y$ axes. The momentum $k_0$ is given by $k_0 = 4 \pi / L_y$, such that two unit cells are present in the simulation box. The box lengths $L_x$ and $L_y$ satisfy $L_x = \frac{\sqrt{3}}{2} L_y$ in order for the triangular solid to be commensurate.

\begin{figure}[t]
\centering
\includegraphics[width=\linewidth]{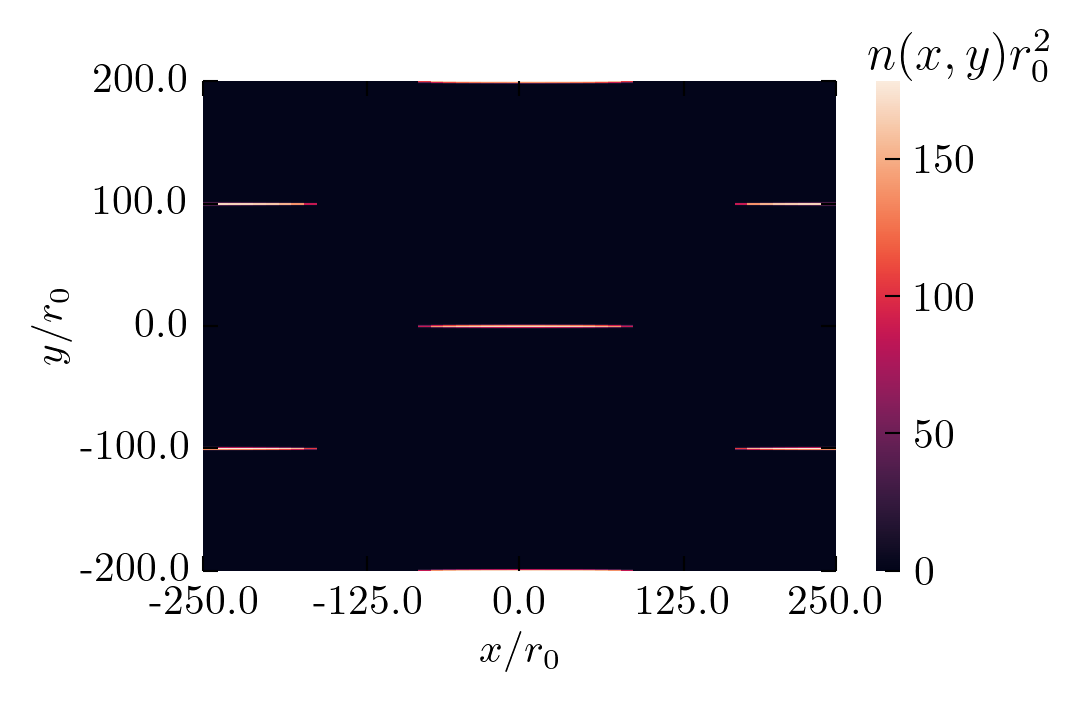}
\caption{Integrated density $n(x,y) = \int dz \abs{\psi({\bf r})}^2$ of the triangular solid solution of Eq.~\ref{eGPE} for $a/a_{\rm dd} = 0.6$, $n_{\rm 2D} r_0^2 = 1$ at $\alpha = 75.9^o$.}
\label{fig_app_2}
\end{figure}

We show the results of our calculations for $\alpha = 75.9^o$ in Table~\ref{table2}, since the other cases follow a similar trend. The reported stripe state energy has been computed for $n_{\rm 2D} r_0^2 = 0.5$, but it stays essentially invariant if the density is decreased as long as the quantity $n_{\rm 2D} L_y$ stays constant. The number of grid points of the simulations have been set to $\{N_x,N_z\} = \{50, 160\}$ points with an imaginary time step of $\Delta \tau = 0.00025 \hbar/E_0$. The number of points along the $y$ axis is set to $N_y = 600$ for $n_{\rm 2D} r_0^2 = 2.5, 5, 10$, $N_y = 1200$ for $n_{\rm 2D} r_0^2 = 0.5, 1$, $N_y = 2400$ for $n_{\rm 2D} r_0^2 = 0.25$ and $N_y = 4800$ for $n_{\rm 2D} r_0^2 = 0.125$. We find that, for all the densities considered, the triangular state features an energy per particle higher than the striped state at $n_{\rm 2D} \rightarrow 0$. In fact, the energy per particle of the triangular state tends to that of the stripe state as the density decreases. This is because the converged triangular solid features extremely anisotropic, striped-like droplets at each lattice position, as shown in Fig.~\ref{fig_app_2}. Notice that, as the density decreases, the box size increases such that the particle number of each droplet yields the optimal energy per particle. As a result, in the limit $n_{\rm 2D} \rightarrow 0$ the optimal triangular solid corresponds to an array of infinitely separated stripes. However, the energy of the triangular solid is always above the striped state due to the extra kinetic energy at the border of the anisotropic droplets. Results at a higher density than the ones considered in Table.~\ref{table2} yield triangular solid states with higher energy per particle than the optimal stripe state until, for high enough density ($n_{\rm 2D} r_0^2 \sim 5$ for $\alpha = 75.9^o$), the elongated droplets merge and the simulation converges instead to a stripe state. Therefore, our results indicate that the stripe state is always energetically favored over the triangular state and, therefore, the latter does not appear in the phase diagram of Fig.~\ref{fig3}.

\bibliography{paper_dipoles_tilt}

\begin{thebibliography}{57}%
\makeatletter
\providecommand \@ifxundefined [1]{%
 \@ifx{#1\undefined}
}%
\providecommand \@ifnum [1]{%
 \ifnum #1\expandafter \@firstoftwo
 \else \expandafter \@secondoftwo
 \fi
}%
\providecommand \@ifx [1]{%
 \ifx #1\expandafter \@firstoftwo
 \else \expandafter \@secondoftwo
 \fi
}%
\providecommand \natexlab [1]{#1}%
\providecommand \enquote  [1]{``#1''}%
\providecommand \bibnamefont  [1]{#1}%
\providecommand \bibfnamefont [1]{#1}%
\providecommand \citenamefont [1]{#1}%
\providecommand \href@noop [0]{\@secondoftwo}%
\providecommand \href [0]{\begingroup \@sanitize@url \@href}%
\providecommand \@href[1]{\@@startlink{#1}\@@href}%
\providecommand \@@href[1]{\endgroup#1\@@endlink}%
\providecommand \@sanitize@url [0]{\catcode `\\12\catcode `\$12\catcode
  `\&12\catcode `\#12\catcode `\^12\catcode `\_12\catcode `\%12\relax}%
\providecommand \@@startlink[1]{}%
\providecommand \@@endlink[0]{}%
\providecommand \url  [0]{\begingroup\@sanitize@url \@url }%
\providecommand \@url [1]{\endgroup\@href {#1}{\urlprefix }}%
\providecommand \urlprefix  [0]{URL }%
\providecommand \Eprint [0]{\href }%
\providecommand \doibase [0]{http://dx.doi.org/}%
\providecommand \selectlanguage [0]{\@gobble}%
\providecommand \bibinfo  [0]{\@secondoftwo}%
\providecommand \bibfield  [0]{\@secondoftwo}%
\providecommand \translation [1]{[#1]}%
\providecommand \BibitemOpen [0]{}%
\providecommand \bibitemStop [0]{}%
\providecommand \bibitemNoStop [0]{.\EOS\space}%
\providecommand \EOS [0]{\spacefactor3000\relax}%
\providecommand \BibitemShut  [1]{\csname bibitem#1\endcsname}%
\let\auto@bib@innerbib\@empty
\bibitem [{\citenamefont {Kadau}\ \emph
  {et~al.}(2016{\natexlab{a}})\citenamefont {Kadau}, \citenamefont {Schmitt},
  \citenamefont {Wenzel}, \citenamefont {Wink}, \citenamefont {Maier},
  \citenamefont {Ferrier-Barbut},\ and\ \citenamefont
  {Pfau}}]{Pfau:nature:2016}%
  \BibitemOpen
  \bibfield  {author} {\bibinfo {author} {\bibfnamefont {Holger}\ \bibnamefont
  {Kadau}}, \bibinfo {author} {\bibfnamefont {Matthias}\ \bibnamefont
  {Schmitt}}, \bibinfo {author} {\bibfnamefont {Matthias}\ \bibnamefont
  {Wenzel}}, \bibinfo {author} {\bibfnamefont {Clarissa}\ \bibnamefont {Wink}},
  \bibinfo {author} {\bibfnamefont {Thomas}\ \bibnamefont {Maier}}, \bibinfo
  {author} {\bibfnamefont {Igor}\ \bibnamefont {Ferrier-Barbut}}, \ and\
  \bibinfo {author} {\bibfnamefont {Tilman}\ \bibnamefont {Pfau}},\ }\bibfield
  {title} {\enquote {\bibinfo {title} {Observing the rosensweig instability of
  a quantum ferrofluid},}\ }\href@noop {} {\bibfield  {journal} {\bibinfo
  {journal} {Nature}\ }\textbf {\bibinfo {volume} {530}},\ \bibinfo {pages}
  {194--197} (\bibinfo {year} {2016}{\natexlab{a}})}\BibitemShut {NoStop}%
\bibitem [{\citenamefont {Schmitt}\ \emph {et~al.}(2016)\citenamefont
  {Schmitt}, \citenamefont {Wenzel}, \citenamefont {B{\"o}ttcher},
  \citenamefont {Ferrier-Barbut},\ and\ \citenamefont
  {Pfau}}]{Pfau:nature2:2016}%
  \BibitemOpen
  \bibfield  {author} {\bibinfo {author} {\bibfnamefont {Matthias}\
  \bibnamefont {Schmitt}}, \bibinfo {author} {\bibfnamefont {Matthias}\
  \bibnamefont {Wenzel}}, \bibinfo {author} {\bibfnamefont {Fabian}\
  \bibnamefont {B{\"o}ttcher}}, \bibinfo {author} {\bibfnamefont {Igor}\
  \bibnamefont {Ferrier-Barbut}}, \ and\ \bibinfo {author} {\bibfnamefont
  {Tilman}\ \bibnamefont {Pfau}},\ }\bibfield  {title} {\enquote {\bibinfo
  {title} {Self-bound droplets of a dilute magnetic quantum liquid},}\
  }\href@noop {} {\bibfield  {journal} {\bibinfo  {journal} {Nature}\ }\textbf
  {\bibinfo {volume} {539}},\ \bibinfo {pages} {259--262} (\bibinfo {year}
  {2016})}\BibitemShut {NoStop}%
\bibitem [{\citenamefont {Ferrier-Barbut}\ \emph {et~al.}(2016)\citenamefont
  {Ferrier-Barbut}, \citenamefont {Kadau}, \citenamefont {Schmitt},
  \citenamefont {Wenzel},\ and\ \citenamefont {Pfau}}]{Pfau:PRL:2016}%
  \BibitemOpen
  \bibfield  {author} {\bibinfo {author} {\bibfnamefont {Igor}\ \bibnamefont
  {Ferrier-Barbut}}, \bibinfo {author} {\bibfnamefont {Holger}\ \bibnamefont
  {Kadau}}, \bibinfo {author} {\bibfnamefont {Matthias}\ \bibnamefont
  {Schmitt}}, \bibinfo {author} {\bibfnamefont {Matthias}\ \bibnamefont
  {Wenzel}}, \ and\ \bibinfo {author} {\bibfnamefont {Tilman}\ \bibnamefont
  {Pfau}},\ }\bibfield  {title} {\enquote {\bibinfo {title} {Observation of
  quantum droplets in a strongly dipolar bose gas},}\ }\href {\doibase
  10.1103/PhysRevLett.116.215301} {\bibfield  {journal} {\bibinfo  {journal}
  {Phys. Rev. Lett.}\ }\textbf {\bibinfo {volume} {116}},\ \bibinfo {pages}
  {215301} (\bibinfo {year} {2016})}\BibitemShut {NoStop}%
\bibitem [{\citenamefont {Chomaz}\ \emph {et~al.}(2016)\citenamefont {Chomaz},
  \citenamefont {Baier}, \citenamefont {Petter}, \citenamefont {Mark},
  \citenamefont {W\"achtler}, \citenamefont {Santos},\ and\ \citenamefont
  {Ferlaino}}]{ferlaino16}%
  \BibitemOpen
  \bibfield  {author} {\bibinfo {author} {\bibfnamefont {L.}~\bibnamefont
  {Chomaz}}, \bibinfo {author} {\bibfnamefont {S.}~\bibnamefont {Baier}},
  \bibinfo {author} {\bibfnamefont {D.}~\bibnamefont {Petter}}, \bibinfo
  {author} {\bibfnamefont {M.~J.}\ \bibnamefont {Mark}}, \bibinfo {author}
  {\bibfnamefont {F.}~\bibnamefont {W\"achtler}}, \bibinfo {author}
  {\bibfnamefont {L.}~\bibnamefont {Santos}}, \ and\ \bibinfo {author}
  {\bibfnamefont {F.}~\bibnamefont {Ferlaino}},\ }\bibfield  {title} {\enquote
  {\bibinfo {title} {Quantum-fluctuation-driven crossover from a dilute
  bose-einstein condensate to a macrodroplet in a dipolar quantum fluid},}\
  }\href {\doibase 10.1103/PhysRevX.6.041039} {\bibfield  {journal} {\bibinfo
  {journal} {Phys. Rev. X}\ }\textbf {\bibinfo {volume} {6}},\ \bibinfo {pages}
  {041039} (\bibinfo {year} {2016})}\BibitemShut {NoStop}%
\bibitem [{\citenamefont {B\"ottcher}\ \emph
  {et~al.}(2019{\natexlab{a}})\citenamefont {B\"ottcher}, \citenamefont
  {Wenzel}, \citenamefont {Schmidt}, \citenamefont {Guo}, \citenamefont
  {Langen}, \citenamefont {Ferrier-Barbut}, \citenamefont {Pfau}, \citenamefont
  {Bomb\'{\i}n}, \citenamefont {S\'anchez-Baena}, \citenamefont {Boronat},\
  and\ \citenamefont {Mazzanti}}]{bottcher19}%
  \BibitemOpen
  \bibfield  {author} {\bibinfo {author} {\bibfnamefont {Fabian}\ \bibnamefont
  {B\"ottcher}}, \bibinfo {author} {\bibfnamefont {Matthias}\ \bibnamefont
  {Wenzel}}, \bibinfo {author} {\bibfnamefont {Jan-Niklas}\ \bibnamefont
  {Schmidt}}, \bibinfo {author} {\bibfnamefont {Mingyang}\ \bibnamefont {Guo}},
  \bibinfo {author} {\bibfnamefont {Tim}\ \bibnamefont {Langen}}, \bibinfo
  {author} {\bibfnamefont {Igor}\ \bibnamefont {Ferrier-Barbut}}, \bibinfo
  {author} {\bibfnamefont {Tilman}\ \bibnamefont {Pfau}}, \bibinfo {author}
  {\bibfnamefont {Ra\'ul}\ \bibnamefont {Bomb\'{\i}n}}, \bibinfo {author}
  {\bibfnamefont {Joan}\ \bibnamefont {S\'anchez-Baena}}, \bibinfo {author}
  {\bibfnamefont {Jordi}\ \bibnamefont {Boronat}}, \ and\ \bibinfo {author}
  {\bibfnamefont {Ferran}\ \bibnamefont {Mazzanti}},\ }\bibfield  {title}
  {\enquote {\bibinfo {title} {Dilute dipolar quantum droplets beyond the
  extended gross-pitaevskii equation},}\ }\href {\doibase
  10.1103/PhysRevResearch.1.033088} {\bibfield  {journal} {\bibinfo  {journal}
  {Phys. Rev. Res.}\ }\textbf {\bibinfo {volume} {1}},\ \bibinfo {pages}
  {033088} (\bibinfo {year} {2019}{\natexlab{a}})}\BibitemShut {NoStop}%
\bibitem [{\citenamefont {Tanzi}\ \emph
  {et~al.}(2019{\natexlab{a}})\citenamefont {Tanzi}, \citenamefont {Lucioni},
  \citenamefont {Fam\`a}, \citenamefont {Catani}, \citenamefont {Fioretti},
  \citenamefont {Gabbanini}, \citenamefont {Bisset}, \citenamefont {Santos},\
  and\ \citenamefont {Modugno}}]{Modugno:PRL:2019}%
  \BibitemOpen
  \bibfield  {author} {\bibinfo {author} {\bibfnamefont {L.}~\bibnamefont
  {Tanzi}}, \bibinfo {author} {\bibfnamefont {E.}~\bibnamefont {Lucioni}},
  \bibinfo {author} {\bibfnamefont {F.}~\bibnamefont {Fam\`a}}, \bibinfo
  {author} {\bibfnamefont {J.}~\bibnamefont {Catani}}, \bibinfo {author}
  {\bibfnamefont {A.}~\bibnamefont {Fioretti}}, \bibinfo {author}
  {\bibfnamefont {C.}~\bibnamefont {Gabbanini}}, \bibinfo {author}
  {\bibfnamefont {R.~N.}\ \bibnamefont {Bisset}}, \bibinfo {author}
  {\bibfnamefont {L.}~\bibnamefont {Santos}}, \ and\ \bibinfo {author}
  {\bibfnamefont {G.}~\bibnamefont {Modugno}},\ }\bibfield  {title} {\enquote
  {\bibinfo {title} {Observation of a dipolar quantum gas with metastable
  supersolid properties},}\ }\href {\doibase 10.1103/PhysRevLett.122.130405}
  {\bibfield  {journal} {\bibinfo  {journal} {Phys. Rev. Lett.}\ }\textbf
  {\bibinfo {volume} {122}},\ \bibinfo {pages} {130405} (\bibinfo {year}
  {2019}{\natexlab{a}})}\BibitemShut {NoStop}%
\bibitem [{\citenamefont {B\"ottcher}\ \emph
  {et~al.}(2019{\natexlab{b}})\citenamefont {B\"ottcher}, \citenamefont
  {Schmidt}, \citenamefont {Wenzel}, \citenamefont {Hertkorn}, \citenamefont
  {Guo}, \citenamefont {Langen},\ and\ \citenamefont {Pfau}}]{Pfau:PRX:2019}%
  \BibitemOpen
  \bibfield  {author} {\bibinfo {author} {\bibfnamefont {Fabian}\ \bibnamefont
  {B\"ottcher}}, \bibinfo {author} {\bibfnamefont {Jan-Niklas}\ \bibnamefont
  {Schmidt}}, \bibinfo {author} {\bibfnamefont {Matthias}\ \bibnamefont
  {Wenzel}}, \bibinfo {author} {\bibfnamefont {Jens}\ \bibnamefont {Hertkorn}},
  \bibinfo {author} {\bibfnamefont {Mingyang}\ \bibnamefont {Guo}}, \bibinfo
  {author} {\bibfnamefont {Tim}\ \bibnamefont {Langen}}, \ and\ \bibinfo
  {author} {\bibfnamefont {Tilman}\ \bibnamefont {Pfau}},\ }\bibfield  {title}
  {\enquote {\bibinfo {title} {Transient supersolid properties in an array of
  dipolar quantum droplets},}\ }\href {\doibase 10.1103/PhysRevX.9.011051}
  {\bibfield  {journal} {\bibinfo  {journal} {Phys. Rev. X}\ }\textbf {\bibinfo
  {volume} {9}},\ \bibinfo {pages} {011051} (\bibinfo {year}
  {2019}{\natexlab{b}})}\BibitemShut {NoStop}%
\bibitem [{\citenamefont {Chomaz}\ \emph {et~al.}(2019)\citenamefont {Chomaz},
  \citenamefont {Petter}, \citenamefont {Ilzh\"ofer}, \citenamefont {Natale},
  \citenamefont {Trautmann}, \citenamefont {Politi}, \citenamefont
  {Durastante}, \citenamefont {van Bijnen}, \citenamefont {Patscheider},
  \citenamefont {Sohmen}, \citenamefont {Mark},\ and\ \citenamefont
  {Ferlaino}}]{Ferlaino:PRX:2019}%
  \BibitemOpen
  \bibfield  {author} {\bibinfo {author} {\bibfnamefont {L.}~\bibnamefont
  {Chomaz}}, \bibinfo {author} {\bibfnamefont {D.}~\bibnamefont {Petter}},
  \bibinfo {author} {\bibfnamefont {P.}~\bibnamefont {Ilzh\"ofer}}, \bibinfo
  {author} {\bibfnamefont {G.}~\bibnamefont {Natale}}, \bibinfo {author}
  {\bibfnamefont {A.}~\bibnamefont {Trautmann}}, \bibinfo {author}
  {\bibfnamefont {C.}~\bibnamefont {Politi}}, \bibinfo {author} {\bibfnamefont
  {G.}~\bibnamefont {Durastante}}, \bibinfo {author} {\bibfnamefont {R.~M.~W.}\
  \bibnamefont {van Bijnen}}, \bibinfo {author} {\bibfnamefont
  {A.}~\bibnamefont {Patscheider}}, \bibinfo {author} {\bibfnamefont
  {M.}~\bibnamefont {Sohmen}}, \bibinfo {author} {\bibfnamefont {M.~J.}\
  \bibnamefont {Mark}}, \ and\ \bibinfo {author} {\bibfnamefont
  {F.}~\bibnamefont {Ferlaino}},\ }\bibfield  {title} {\enquote {\bibinfo
  {title} {Long-lived and transient supersolid behaviors in dipolar quantum
  gases},}\ }\href {\doibase 10.1103/PhysRevX.9.021012} {\bibfield  {journal}
  {\bibinfo  {journal} {Phys. Rev. X}\ }\textbf {\bibinfo {volume} {9}},\
  \bibinfo {pages} {021012} (\bibinfo {year} {2019})}\BibitemShut {NoStop}%
\bibitem [{\citenamefont {Tanzi}\ \emph
  {et~al.}(2019{\natexlab{b}})\citenamefont {Tanzi}, \citenamefont {Roccuzzo},
  \citenamefont {Lucioni}, \citenamefont {Fam{\`a}}, \citenamefont {Fioretti},
  \citenamefont {Gabbanini}, \citenamefont {Modugno}, \citenamefont {Recati},\
  and\ \citenamefont {Stringari}}]{Tanzi:Nature:2019}%
  \BibitemOpen
  \bibfield  {author} {\bibinfo {author} {\bibfnamefont {L.}~\bibnamefont
  {Tanzi}}, \bibinfo {author} {\bibfnamefont {S.~M.}\ \bibnamefont {Roccuzzo}},
  \bibinfo {author} {\bibfnamefont {E.}~\bibnamefont {Lucioni}}, \bibinfo
  {author} {\bibfnamefont {F.}~\bibnamefont {Fam{\`a}}}, \bibinfo {author}
  {\bibfnamefont {A.}~\bibnamefont {Fioretti}}, \bibinfo {author}
  {\bibfnamefont {C.}~\bibnamefont {Gabbanini}}, \bibinfo {author}
  {\bibfnamefont {G.}~\bibnamefont {Modugno}}, \bibinfo {author} {\bibfnamefont
  {A.}~\bibnamefont {Recati}}, \ and\ \bibinfo {author} {\bibfnamefont
  {S.}~\bibnamefont {Stringari}},\ }\bibfield  {title} {\enquote {\bibinfo
  {title} {Supersolid symmetry breaking from compressional oscillations in a
  dipolar quantum gas},}\ }\href {\doibase 10.1038/s41586-019-1568-6}
  {\bibfield  {journal} {\bibinfo  {journal} {Nature}\ }\textbf {\bibinfo
  {volume} {574}},\ \bibinfo {pages} {382--385} (\bibinfo {year}
  {2019}{\natexlab{b}})}\BibitemShut {NoStop}%
\bibitem [{\citenamefont {Guo}\ \emph {et~al.}(2019)\citenamefont {Guo},
  \citenamefont {B{\"o}ttcher}, \citenamefont {Hertkorn}, \citenamefont
  {Schmidt}, \citenamefont {Wenzel}, \citenamefont {B{\"u}chler}, \citenamefont
  {Langen},\ and\ \citenamefont {Pfau}}]{Guo:Nature:2019}%
  \BibitemOpen
  \bibfield  {author} {\bibinfo {author} {\bibfnamefont {Mingyang}\
  \bibnamefont {Guo}}, \bibinfo {author} {\bibfnamefont {Fabian}\ \bibnamefont
  {B{\"o}ttcher}}, \bibinfo {author} {\bibfnamefont {Jens}\ \bibnamefont
  {Hertkorn}}, \bibinfo {author} {\bibfnamefont {Jan-Niklas}\ \bibnamefont
  {Schmidt}}, \bibinfo {author} {\bibfnamefont {Matthias}\ \bibnamefont
  {Wenzel}}, \bibinfo {author} {\bibfnamefont {Hans~Peter}\ \bibnamefont
  {B{\"u}chler}}, \bibinfo {author} {\bibfnamefont {Tim}\ \bibnamefont
  {Langen}}, \ and\ \bibinfo {author} {\bibfnamefont {Tilman}\ \bibnamefont
  {Pfau}},\ }\bibfield  {title} {\enquote {\bibinfo {title} {The low-energy
  goldstone mode in a trapped dipolar supersolid},}\ }\href {\doibase
  10.1038/s41586-019-1569-5} {\bibfield  {journal} {\bibinfo  {journal}
  {Nature}\ }\textbf {\bibinfo {volume} {574}},\ \bibinfo {pages} {386--389}
  (\bibinfo {year} {2019})}\BibitemShut {NoStop}%
\bibitem [{\citenamefont {Tanzi}\ \emph {et~al.}(2021)\citenamefont {Tanzi},
  \citenamefont {Maloberti}, \citenamefont {Biagioni}, \citenamefont
  {Fioretti}, \citenamefont {Gabbanini},\ and\ \citenamefont
  {Modugno}}]{Tanzi:Science:2021}%
  \BibitemOpen
  \bibfield  {author} {\bibinfo {author} {\bibfnamefont {L.}~\bibnamefont
  {Tanzi}}, \bibinfo {author} {\bibfnamefont {J.~G.}\ \bibnamefont
  {Maloberti}}, \bibinfo {author} {\bibfnamefont {G.}~\bibnamefont {Biagioni}},
  \bibinfo {author} {\bibfnamefont {A.}~\bibnamefont {Fioretti}}, \bibinfo
  {author} {\bibfnamefont {C.}~\bibnamefont {Gabbanini}}, \ and\ \bibinfo
  {author} {\bibfnamefont {G.}~\bibnamefont {Modugno}},\ }\bibfield  {title}
  {\enquote {\bibinfo {title} {Evidence of superfluidity in a dipolar
  supersolid from nonclassical rotational inertia},}\ }\href {\doibase
  10.1126/science.aba4309} {\bibfield  {journal} {\bibinfo  {journal}
  {Science}\ }\textbf {\bibinfo {volume} {371}},\ \bibinfo {pages} {1162--1165}
  (\bibinfo {year} {2021})}\BibitemShut {NoStop}%
\bibitem [{\citenamefont {Norcia}\ \emph {et~al.}(2021)\citenamefont {Norcia},
  \citenamefont {Politi}, \citenamefont {Klaus}, \citenamefont {Poli},
  \citenamefont {Sohmen}, \citenamefont {Mark}, \citenamefont {Bisset},
  \citenamefont {Santos},\ and\ \citenamefont {Ferlaino}}]{norcia21:nature}%
  \BibitemOpen
  \bibfield  {author} {\bibinfo {author} {\bibfnamefont {Matthew~A.}\
  \bibnamefont {Norcia}}, \bibinfo {author} {\bibfnamefont {Claudia}\
  \bibnamefont {Politi}}, \bibinfo {author} {\bibfnamefont {Lauritz}\
  \bibnamefont {Klaus}}, \bibinfo {author} {\bibfnamefont {Elena}\ \bibnamefont
  {Poli}}, \bibinfo {author} {\bibfnamefont {Maximilian}\ \bibnamefont
  {Sohmen}}, \bibinfo {author} {\bibfnamefont {Manfred~J.}\ \bibnamefont
  {Mark}}, \bibinfo {author} {\bibfnamefont {Russell~N.}\ \bibnamefont
  {Bisset}}, \bibinfo {author} {\bibfnamefont {Luis}\ \bibnamefont {Santos}}, \
  and\ \bibinfo {author} {\bibfnamefont {Francesca}\ \bibnamefont {Ferlaino}},\
  }\bibfield  {title} {\enquote {\bibinfo {title} {Two-dimensional
  supersolidity in a dipolar quantum gas},}\ }\href {\doibase
  10.1038/s41586-021-03725-7} {\bibfield  {journal} {\bibinfo  {journal}
  {Nature}\ }\textbf {\bibinfo {volume} {596}},\ \bibinfo {pages} {357--361}
  (\bibinfo {year} {2021})}\BibitemShut {NoStop}%
\bibitem [{\citenamefont {Biagioni}\ \emph {et~al.}(2022)\citenamefont
  {Biagioni}, \citenamefont {Antolini}, \citenamefont {Ala\~na}, \citenamefont
  {Modugno}, \citenamefont {Fioretti}, \citenamefont {Gabbanini}, \citenamefont
  {Tanzi},\ and\ \citenamefont {Modugno}}]{BiagioniPRX2022}%
  \BibitemOpen
  \bibfield  {author} {\bibinfo {author} {\bibfnamefont {Giulio}\ \bibnamefont
  {Biagioni}}, \bibinfo {author} {\bibfnamefont {Nicol\`o}\ \bibnamefont
  {Antolini}}, \bibinfo {author} {\bibfnamefont {Aitor}\ \bibnamefont
  {Ala\~na}}, \bibinfo {author} {\bibfnamefont {Michele}\ \bibnamefont
  {Modugno}}, \bibinfo {author} {\bibfnamefont {Andrea}\ \bibnamefont
  {Fioretti}}, \bibinfo {author} {\bibfnamefont {Carlo}\ \bibnamefont
  {Gabbanini}}, \bibinfo {author} {\bibfnamefont {Luca}\ \bibnamefont {Tanzi}},
  \ and\ \bibinfo {author} {\bibfnamefont {Giovanni}\ \bibnamefont {Modugno}},\
  }\bibfield  {title} {\enquote {\bibinfo {title} {Dimensional crossover in the
  superfluid-supersolid quantum phase transition},}\ }\href {\doibase
  10.1103/PhysRevX.12.021019} {\bibfield  {journal} {\bibinfo  {journal} {Phys.
  Rev. X}\ }\textbf {\bibinfo {volume} {12}},\ \bibinfo {pages} {021019}
  (\bibinfo {year} {2022})}\BibitemShut {NoStop}%
\bibitem [{\citenamefont {Sohmen}\ \emph {et~al.}(2021)\citenamefont {Sohmen},
  \citenamefont {Politi}, \citenamefont {Klaus}, \citenamefont {Chomaz},
  \citenamefont {Mark}, \citenamefont {Norcia},\ and\ \citenamefont
  {Ferlaino}}]{Ferlaino:PRL:2021}%
  \BibitemOpen
  \bibfield  {author} {\bibinfo {author} {\bibfnamefont {Maximilian}\
  \bibnamefont {Sohmen}}, \bibinfo {author} {\bibfnamefont {Claudia}\
  \bibnamefont {Politi}}, \bibinfo {author} {\bibfnamefont {Lauritz}\
  \bibnamefont {Klaus}}, \bibinfo {author} {\bibfnamefont {Lauriane}\
  \bibnamefont {Chomaz}}, \bibinfo {author} {\bibfnamefont {Manfred~J.}\
  \bibnamefont {Mark}}, \bibinfo {author} {\bibfnamefont {Matthew~A.}\
  \bibnamefont {Norcia}}, \ and\ \bibinfo {author} {\bibfnamefont {Francesca}\
  \bibnamefont {Ferlaino}},\ }\bibfield  {title} {\enquote {\bibinfo {title}
  {Birth, life, and death of a dipolar supersolid},}\ }\href {\doibase
  10.1103/PhysRevLett.126.233401} {\bibfield  {journal} {\bibinfo  {journal}
  {Phys. Rev. Lett.}\ }\textbf {\bibinfo {volume} {126}},\ \bibinfo {pages}
  {233401} (\bibinfo {year} {2021})}\BibitemShut {NoStop}%
\bibitem [{\citenamefont {S{\'a}nchez-Baena}\ \emph {et~al.}(2023)\citenamefont
  {S{\'a}nchez-Baena}, \citenamefont {Politi}, \citenamefont {Maucher},
  \citenamefont {Ferlaino},\ and\ \citenamefont {Pohl}}]{baena22}%
  \BibitemOpen
  \bibfield  {author} {\bibinfo {author} {\bibfnamefont {J.}~\bibnamefont
  {S{\'a}nchez-Baena}}, \bibinfo {author} {\bibfnamefont {C.}~\bibnamefont
  {Politi}}, \bibinfo {author} {\bibfnamefont {F.}~\bibnamefont {Maucher}},
  \bibinfo {author} {\bibfnamefont {F.}~\bibnamefont {Ferlaino}}, \ and\
  \bibinfo {author} {\bibfnamefont {T.}~\bibnamefont {Pohl}},\ }\bibfield
  {title} {\enquote {\bibinfo {title} {Heating a dipolar quantum fluid into a
  solid},}\ }\href {\doibase 10.1038/s41467-023-37207-3} {\bibfield  {journal}
  {\bibinfo  {journal} {Nature Communications}\ }\textbf {\bibinfo {volume}
  {14}},\ \bibinfo {pages} {1868} (\bibinfo {year} {2023})}\BibitemShut
  {NoStop}%
\bibitem [{\citenamefont {S\'anchez-Baena}\ \emph {et~al.}(2024)\citenamefont
  {S\'anchez-Baena}, \citenamefont {Pohl},\ and\ \citenamefont
  {Maucher}}]{baena24}%
  \BibitemOpen
  \bibfield  {author} {\bibinfo {author} {\bibfnamefont {J.}~\bibnamefont
  {S\'anchez-Baena}}, \bibinfo {author} {\bibfnamefont {T.}~\bibnamefont
  {Pohl}}, \ and\ \bibinfo {author} {\bibfnamefont {F.}~\bibnamefont
  {Maucher}},\ }\bibfield  {title} {\enquote {\bibinfo {title}
  {Superfluid-supersolid phase transition of elongated dipolar bose-einstein
  condensates at finite temperatures},}\ }\href {\doibase
  10.1103/PhysRevResearch.6.023183} {\bibfield  {journal} {\bibinfo  {journal}
  {Phys. Rev. Res.}\ }\textbf {\bibinfo {volume} {6}},\ \bibinfo {pages}
  {023183} (\bibinfo {year} {2024})}\BibitemShut {NoStop}%
\bibitem [{\citenamefont {He}\ \emph {et~al.}(2024{\natexlab{a}})\citenamefont
  {He}, \citenamefont {Sanchez-Baena}, \citenamefont {Maucher},\ and\
  \citenamefont {Zhang}}]{He2024}%
  \BibitemOpen
  \bibfield  {author} {\bibinfo {author} {\bibfnamefont {Liang-Jun}\
  \bibnamefont {He}}, \bibinfo {author} {\bibfnamefont {Juan}\ \bibnamefont
  {Sanchez-Baena}}, \bibinfo {author} {\bibfnamefont {Fabian}\ \bibnamefont
  {Maucher}}, \ and\ \bibinfo {author} {\bibfnamefont {Yong-Chang}\
  \bibnamefont {Zhang}},\ }\href {https://arxiv.org/abs/2410.19260} {\enquote
  {\bibinfo {title} {Accessing elusive two-dimensional phases of dipolar
  bose-einstein condensates by finite temperature},}\ } (\bibinfo {year}
  {2024}{\natexlab{a}}),\ \Eprint {http://arxiv.org/abs/2410.19260}
  {arXiv:2410.19260 [cond-mat.quant-gas]} \BibitemShut {NoStop}%
\bibitem [{\citenamefont {Sunami}\ \emph {et~al.}(2022)\citenamefont {Sunami},
  \citenamefont {Singh}, \citenamefont {Garrick}, \citenamefont {Beregi},
  \citenamefont {Barker}, \citenamefont {Luksch}, \citenamefont {Bentine},
  \citenamefont {Mathey},\ and\ \citenamefont {Foot}}]{sunami2022:prl}%
  \BibitemOpen
  \bibfield  {author} {\bibinfo {author} {\bibfnamefont {S.}~\bibnamefont
  {Sunami}}, \bibinfo {author} {\bibfnamefont {V.~P.}\ \bibnamefont {Singh}},
  \bibinfo {author} {\bibfnamefont {D.}~\bibnamefont {Garrick}}, \bibinfo
  {author} {\bibfnamefont {A.}~\bibnamefont {Beregi}}, \bibinfo {author}
  {\bibfnamefont {A.~J.}\ \bibnamefont {Barker}}, \bibinfo {author}
  {\bibfnamefont {K.}~\bibnamefont {Luksch}}, \bibinfo {author} {\bibfnamefont
  {E.}~\bibnamefont {Bentine}}, \bibinfo {author} {\bibfnamefont
  {L.}~\bibnamefont {Mathey}}, \ and\ \bibinfo {author} {\bibfnamefont {C.~J.}\
  \bibnamefont {Foot}},\ }\bibfield  {title} {\enquote {\bibinfo {title}
  {Observation of the berezinskii-kosterlitz-thouless transition in a
  two-dimensional bose gas via matter-wave interferometry},}\ }\href {\doibase
  10.1103/PhysRevLett.128.250402} {\bibfield  {journal} {\bibinfo  {journal}
  {Phys. Rev. Lett.}\ }\textbf {\bibinfo {volume} {128}},\ \bibinfo {pages}
  {250402} (\bibinfo {year} {2022})}\BibitemShut {NoStop}%
\bibitem [{\citenamefont {Yu}\ \emph {et~al.}(2024)\citenamefont {Yu},
  \citenamefont {Bhave}, \citenamefont {Reeve}, \citenamefont {Song},\ and\
  \citenamefont {Schneider}}]{Yu2024:science}%
  \BibitemOpen
  \bibfield  {author} {\bibinfo {author} {\bibfnamefont {Jr-Chiun}\
  \bibnamefont {Yu}}, \bibinfo {author} {\bibfnamefont {Shaurya}\ \bibnamefont
  {Bhave}}, \bibinfo {author} {\bibfnamefont {Lee}\ \bibnamefont {Reeve}},
  \bibinfo {author} {\bibfnamefont {Bo}~\bibnamefont {Song}}, \ and\ \bibinfo
  {author} {\bibfnamefont {Ulrich}\ \bibnamefont {Schneider}},\ }\bibfield
  {title} {\enquote {\bibinfo {title} {Observing the two-dimensional bose glass
  in an optical quasicrystal},}\ }\href {\doibase 10.1038/s41586-024-07875-2}
  {\bibfield  {journal} {\bibinfo  {journal} {Nature}\ }\textbf {\bibinfo
  {volume} {633}},\ \bibinfo {pages} {338--343} (\bibinfo {year}
  {2024})}\BibitemShut {NoStop}%
\bibitem [{\citenamefont {Guo}\ \emph {et~al.}(2024)\citenamefont {Guo},
  \citenamefont {Yao}, \citenamefont {Ramanjanappa}, \citenamefont {Dhar},
  \citenamefont {Horvath}, \citenamefont {Pizzino}, \citenamefont {Giamarchi},
  \citenamefont {Landini},\ and\ \citenamefont {N{\"a}gerl}}]{Guo2024:science}%
  \BibitemOpen
  \bibfield  {author} {\bibinfo {author} {\bibfnamefont {Yanliang}\
  \bibnamefont {Guo}}, \bibinfo {author} {\bibfnamefont {Hepeng}\ \bibnamefont
  {Yao}}, \bibinfo {author} {\bibfnamefont {Satwik}\ \bibnamefont
  {Ramanjanappa}}, \bibinfo {author} {\bibfnamefont {Sudipta}\ \bibnamefont
  {Dhar}}, \bibinfo {author} {\bibfnamefont {Milena}\ \bibnamefont {Horvath}},
  \bibinfo {author} {\bibfnamefont {Lorenzo}\ \bibnamefont {Pizzino}}, \bibinfo
  {author} {\bibfnamefont {Thierry}\ \bibnamefont {Giamarchi}}, \bibinfo
  {author} {\bibfnamefont {Manuele}\ \bibnamefont {Landini}}, \ and\ \bibinfo
  {author} {\bibfnamefont {Hanns-Christoph}\ \bibnamefont {N{\"a}gerl}},\
  }\bibfield  {title} {\enquote {\bibinfo {title} {Observation of the 2d--1d
  crossover in strongly interacting ultracold bosons},}\ }\href {\doibase
  10.1038/s41567-024-02459-3} {\bibfield  {journal} {\bibinfo  {journal}
  {Nature Physics}\ }\textbf {\bibinfo {volume} {20}},\ \bibinfo {pages}
  {934--938} (\bibinfo {year} {2024})}\BibitemShut {NoStop}%
\bibitem [{\citenamefont {Lima}\ and\ \citenamefont
  {Pelster}(2011)}]{Lima:2011eq}%
  \BibitemOpen
  \bibfield  {author} {\bibinfo {author} {\bibfnamefont {Aristeu R~P}\
  \bibnamefont {Lima}}\ and\ \bibinfo {author} {\bibfnamefont {Axel}\
  \bibnamefont {Pelster}},\ }\bibfield  {title} {\enquote {\bibinfo {title}
  {{Quantum fluctuations in dipolar Bose gases}},}\ }\href@noop {} {\bibfield
  {journal} {\bibinfo  {journal} {Physical Review A}\ }\textbf {\bibinfo
  {volume} {84}},\ \bibinfo {pages} {041604--4} (\bibinfo {year}
  {2011})}\BibitemShut {NoStop}%
\bibitem [{\citenamefont {Lima}\ and\ \citenamefont
  {Pelster}(2012)}]{pelster12}%
  \BibitemOpen
  \bibfield  {author} {\bibinfo {author} {\bibfnamefont {A.~R.~P.}\
  \bibnamefont {Lima}}\ and\ \bibinfo {author} {\bibfnamefont {A.}~\bibnamefont
  {Pelster}},\ }\bibfield  {title} {\enquote {\bibinfo {title} {Beyond
  mean-field low-lying excitations of dipolar bose gases},}\ }\href {\doibase
  10.1103/PhysRevA.86.063609} {\bibfield  {journal} {\bibinfo  {journal} {Phys.
  Rev. A}\ }\textbf {\bibinfo {volume} {86}},\ \bibinfo {pages} {063609}
  (\bibinfo {year} {2012})}\BibitemShut {NoStop}%
\bibitem [{\citenamefont {W{\"a}chtler}\ and\ \citenamefont
  {Santos}(2016)}]{Wachtler:2016kk}%
  \BibitemOpen
  \bibfield  {author} {\bibinfo {author} {\bibfnamefont {F}~\bibnamefont
  {W{\"a}chtler}}\ and\ \bibinfo {author} {\bibfnamefont {L}~\bibnamefont
  {Santos}},\ }\bibfield  {title} {\enquote {\bibinfo {title} {{Ground-state
  properties and elementary excitations of quantum droplets in dipolar
  Bose-Einstein condensates}},}\ }\href@noop {} {\bibfield  {journal} {\bibinfo
   {journal} {Physical Review A}\ }\textbf {\bibinfo {volume} {94}},\ \bibinfo
  {pages} {043618--7} (\bibinfo {year} {2016})}\BibitemShut {NoStop}%
\bibitem [{\citenamefont {Edler}\ \emph {et~al.}(2017)\citenamefont {Edler},
  \citenamefont {Mishra}, \citenamefont {W\"achtler}, \citenamefont {Nath},
  \citenamefont {Sinha},\ and\ \citenamefont {Santos}}]{edler_PRL_2016}%
  \BibitemOpen
  \bibfield  {author} {\bibinfo {author} {\bibfnamefont {D.}~\bibnamefont
  {Edler}}, \bibinfo {author} {\bibfnamefont {C.}~\bibnamefont {Mishra}},
  \bibinfo {author} {\bibfnamefont {F.}~\bibnamefont {W\"achtler}}, \bibinfo
  {author} {\bibfnamefont {R.}~\bibnamefont {Nath}}, \bibinfo {author}
  {\bibfnamefont {S.}~\bibnamefont {Sinha}}, \ and\ \bibinfo {author}
  {\bibfnamefont {L.}~\bibnamefont {Santos}},\ }\bibfield  {title} {\enquote
  {\bibinfo {title} {Quantum fluctuations in quasi-one-dimensional dipolar
  bose-einstein condensates},}\ }\href {\doibase
  10.1103/PhysRevLett.119.050403} {\bibfield  {journal} {\bibinfo  {journal}
  {Phys. Rev. Lett.}\ }\textbf {\bibinfo {volume} {119}},\ \bibinfo {pages}
  {050403} (\bibinfo {year} {2017})}\BibitemShut {NoStop}%
\bibitem [{\citenamefont {Zin}\ \emph {et~al.}(2021)\citenamefont {Zin},
  \citenamefont {Pylak}, \citenamefont {Wasak}, \citenamefont {Jachymski},\
  and\ \citenamefont {Idziaszek}}]{Zin_2021}%
  \BibitemOpen
  \bibfield  {author} {\bibinfo {author} {\bibfnamefont {Paweł}\ \bibnamefont
  {Zin}}, \bibinfo {author} {\bibfnamefont {Maciej}\ \bibnamefont {Pylak}},
  \bibinfo {author} {\bibfnamefont {Tomasz}\ \bibnamefont {Wasak}}, \bibinfo
  {author} {\bibfnamefont {Krzysztof}\ \bibnamefont {Jachymski}}, \ and\
  \bibinfo {author} {\bibfnamefont {Zbigniew}\ \bibnamefont {Idziaszek}},\
  }\bibfield  {title} {\enquote {\bibinfo {title} {Quantum droplets in a
  dipolar bose gas at a dimensional crossover},}\ }\href {\doibase
  10.1088/1361-6455/ac2244} {\bibfield  {journal} {\bibinfo  {journal} {Journal
  of Physics B: Atomic, Molecular and Optical Physics}\ }\textbf {\bibinfo
  {volume} {54}},\ \bibinfo {pages} {165302} (\bibinfo {year}
  {2021})}\BibitemShut {NoStop}%
\bibitem [{\citenamefont {Du}\ \emph {et~al.}(2024)\citenamefont {Du},
  \citenamefont {Barral}, \citenamefont {Cantara}, \citenamefont {de~Hond},
  \citenamefont {Lu},\ and\ \citenamefont {Ketterle}}]{li_du2024:science}%
  \BibitemOpen
  \bibfield  {author} {\bibinfo {author} {\bibfnamefont {Li}~\bibnamefont
  {Du}}, \bibinfo {author} {\bibfnamefont {Pierre}\ \bibnamefont {Barral}},
  \bibinfo {author} {\bibfnamefont {Michael}\ \bibnamefont {Cantara}}, \bibinfo
  {author} {\bibfnamefont {Julius}\ \bibnamefont {de~Hond}}, \bibinfo {author}
  {\bibfnamefont {Yu-Kun}\ \bibnamefont {Lu}}, \ and\ \bibinfo {author}
  {\bibfnamefont {Wolfgang}\ \bibnamefont {Ketterle}},\ }\bibfield  {title}
  {\enquote {\bibinfo {title} {Atomic physics on a 50-nm scale: Realization of
  a bilayer system of dipolar atoms},}\ }\href {\doibase
  10.1126/science.adh3023} {\bibfield  {journal} {\bibinfo  {journal}
  {Science}\ }\textbf {\bibinfo {volume} {384}},\ \bibinfo {pages} {546--551}
  (\bibinfo {year} {2024})},\ \Eprint
  {http://arxiv.org/abs/https://www.science.org/doi/pdf/10.1126/science.adh3023}
  {https://www.science.org/doi/pdf/10.1126/science.adh3023} \BibitemShut
  {NoStop}%
\bibitem [{\citenamefont {He}\ \emph {et~al.}(2024{\natexlab{b}})\citenamefont
  {He}, \citenamefont {Chen}, \citenamefont {Zhen}, \citenamefont {Huang},
  \citenamefont {Parit},\ and\ \citenamefont {Jo}}]{he2024:arxiv}%
  \BibitemOpen
  \bibfield  {author} {\bibinfo {author} {\bibfnamefont {Yifei}\ \bibnamefont
  {He}}, \bibinfo {author} {\bibfnamefont {Ziting}\ \bibnamefont {Chen}},
  \bibinfo {author} {\bibfnamefont {Haoting}\ \bibnamefont {Zhen}}, \bibinfo
  {author} {\bibfnamefont {Mingchen}\ \bibnamefont {Huang}}, \bibinfo {author}
  {\bibfnamefont {Mithilesh~K}\ \bibnamefont {Parit}}, \ and\ \bibinfo {author}
  {\bibfnamefont {Gyu-Boong}\ \bibnamefont {Jo}},\ }\href
  {https://arxiv.org/abs/2403.18683} {\enquote {\bibinfo {title} {Exploring the
  berezinskii-kosterlitz-thouless transition in a two-dimensional dipolar bose
  gas},}\ } (\bibinfo {year} {2024}{\natexlab{b}}),\ \Eprint
  {http://arxiv.org/abs/2403.18683} {arXiv:2403.18683 [cond-mat.quant-gas]}
  \BibitemShut {NoStop}%
\bibitem [{\citenamefont {Fischer}(2006)}]{fischer_06_PRA}%
  \BibitemOpen
  \bibfield  {author} {\bibinfo {author} {\bibfnamefont {Uwe~R.}\ \bibnamefont
  {Fischer}},\ }\bibfield  {title} {\enquote {\bibinfo {title} {Stability of
  quasi-two-dimensional bose-einstein condensates with dominant dipole-dipole
  interactions},}\ }\href {\doibase 10.1103/PhysRevA.73.031602} {\bibfield
  {journal} {\bibinfo  {journal} {Phys. Rev. A}\ }\textbf {\bibinfo {volume}
  {73}},\ \bibinfo {pages} {031602} (\bibinfo {year} {2006})}\BibitemShut
  {NoStop}%
\bibitem [{\citenamefont {Boudjem\^aa}\ and\ \citenamefont
  {Shlyapnikov}(2013)}]{boudjemaa_2013_PRA}%
  \BibitemOpen
  \bibfield  {author} {\bibinfo {author} {\bibfnamefont {Abdel\^aali}\
  \bibnamefont {Boudjem\^aa}}\ and\ \bibinfo {author} {\bibfnamefont {G.~V.}\
  \bibnamefont {Shlyapnikov}},\ }\bibfield  {title} {\enquote {\bibinfo {title}
  {Two-dimensional dipolar bose gas with the roton-maxon excitation
  spectrum},}\ }\href {\doibase 10.1103/PhysRevA.87.025601} {\bibfield
  {journal} {\bibinfo  {journal} {Phys. Rev. A}\ }\textbf {\bibinfo {volume}
  {87}},\ \bibinfo {pages} {025601} (\bibinfo {year} {2013})}\BibitemShut
  {NoStop}%
\bibitem [{\citenamefont {Mishra}\ and\ \citenamefont
  {Nath}(2016)}]{mishra_2016_PRA}%
  \BibitemOpen
  \bibfield  {author} {\bibinfo {author} {\bibfnamefont {Chinmayee}\
  \bibnamefont {Mishra}}\ and\ \bibinfo {author} {\bibfnamefont {Rejish}\
  \bibnamefont {Nath}},\ }\bibfield  {title} {\enquote {\bibinfo {title}
  {Dipolar condensates with tilted dipoles in a pancake-shaped confinement},}\
  }\href {\doibase 10.1103/PhysRevA.94.033633} {\bibfield  {journal} {\bibinfo
  {journal} {Phys. Rev. A}\ }\textbf {\bibinfo {volume} {94}},\ \bibinfo
  {pages} {033633} (\bibinfo {year} {2016})}\BibitemShut {NoStop}%
\bibitem [{\citenamefont {Baillie}\ and\ \citenamefont
  {Blakie}(2015)}]{Baillie_2015_NJP}%
  \BibitemOpen
  \bibfield  {author} {\bibinfo {author} {\bibfnamefont {D}~\bibnamefont
  {Baillie}}\ and\ \bibinfo {author} {\bibfnamefont {P~B}\ \bibnamefont
  {Blakie}},\ }\bibfield  {title} {\enquote {\bibinfo {title} {A general theory
  of flattened dipolar condensates},}\ }\href {\doibase
  10.1088/1367-2630/17/3/033028} {\bibfield  {journal} {\bibinfo  {journal}
  {New Journal of Physics}\ }\textbf {\bibinfo {volume} {17}},\ \bibinfo
  {pages} {033028} (\bibinfo {year} {2015})}\BibitemShut {NoStop}%
\bibitem [{\citenamefont {Shen}\ and\ \citenamefont
  {Quader}(2021)}]{pengtao2021}%
  \BibitemOpen
  \bibfield  {author} {\bibinfo {author} {\bibfnamefont {Pengtao}\ \bibnamefont
  {Shen}}\ and\ \bibinfo {author} {\bibfnamefont {Khandker~F.}\ \bibnamefont
  {Quader}},\ }\bibfield  {title} {\enquote {\bibinfo {title}
  {Finite-temperature instabilities of a two-dimensional dipolar bose gas at
  arbitrary tilt angle},}\ }\href {\doibase 10.1103/PhysRevA.103.043317}
  {\bibfield  {journal} {\bibinfo  {journal} {Phys. Rev. A}\ }\textbf {\bibinfo
  {volume} {103}},\ \bibinfo {pages} {043317} (\bibinfo {year}
  {2021})}\BibitemShut {NoStop}%
\bibitem [{\citenamefont {Fedorov}\ \emph {et~al.}(2014)\citenamefont
  {Fedorov}, \citenamefont {Kurbakov}, \citenamefont {Shchadilova},\ and\
  \citenamefont {Lozovik}}]{fedorov_2014_PRA}%
  \BibitemOpen
  \bibfield  {author} {\bibinfo {author} {\bibfnamefont {A.~K.}\ \bibnamefont
  {Fedorov}}, \bibinfo {author} {\bibfnamefont {I.~L.}\ \bibnamefont
  {Kurbakov}}, \bibinfo {author} {\bibfnamefont {Y.~E.}\ \bibnamefont
  {Shchadilova}}, \ and\ \bibinfo {author} {\bibfnamefont {Yu.~E.}\
  \bibnamefont {Lozovik}},\ }\bibfield  {title} {\enquote {\bibinfo {title}
  {Two-dimensional bose gas of tilted dipoles: Roton instability and condensate
  depletion},}\ }\href {\doibase 10.1103/PhysRevA.90.043616} {\bibfield
  {journal} {\bibinfo  {journal} {Phys. Rev. A}\ }\textbf {\bibinfo {volume}
  {90}},\ \bibinfo {pages} {043616} (\bibinfo {year} {2014})}\BibitemShut
  {NoStop}%
\bibitem [{\citenamefont {Aleksandrova}\ \emph {et~al.}(2024)\citenamefont
  {Aleksandrova}, \citenamefont {Kurbakov}, \citenamefont {Fedorov},\ and\
  \citenamefont {Lozovik}}]{aleksandrova_2024_PRA}%
  \BibitemOpen
  \bibfield  {author} {\bibinfo {author} {\bibfnamefont {A.~N.}\ \bibnamefont
  {Aleksandrova}}, \bibinfo {author} {\bibfnamefont {I.~L.}\ \bibnamefont
  {Kurbakov}}, \bibinfo {author} {\bibfnamefont {A.~K.}\ \bibnamefont
  {Fedorov}}, \ and\ \bibinfo {author} {\bibfnamefont {Yu.~E.}\ \bibnamefont
  {Lozovik}},\ }\bibfield  {title} {\enquote {\bibinfo {title}
  {Density-wave-type supersolid of two-dimensional tilted dipolar bosons},}\
  }\href {\doibase 10.1103/PhysRevA.109.063326} {\bibfield  {journal} {\bibinfo
   {journal} {Phys. Rev. A}\ }\textbf {\bibinfo {volume} {109}},\ \bibinfo
  {pages} {063326} (\bibinfo {year} {2024})}\BibitemShut {NoStop}%
\bibitem [{\citenamefont {Marchetti}\ and\ \citenamefont
  {Parish}(2013)}]{marchetti2013:prb}%
  \BibitemOpen
  \bibfield  {author} {\bibinfo {author} {\bibfnamefont {F.~M.}\ \bibnamefont
  {Marchetti}}\ and\ \bibinfo {author} {\bibfnamefont {M.~M.}\ \bibnamefont
  {Parish}},\ }\bibfield  {title} {\enquote {\bibinfo {title} {Density-wave
  phases of dipolar fermions in a bilayer},}\ }\href {\doibase
  10.1103/PhysRevB.87.045110} {\bibfield  {journal} {\bibinfo  {journal} {Phys.
  Rev. B}\ }\textbf {\bibinfo {volume} {87}},\ \bibinfo {pages} {045110}
  (\bibinfo {year} {2013})}\BibitemShut {NoStop}%
\bibitem [{\citenamefont {Block}\ and\ \citenamefont
  {Bruun}(2014)}]{block2014:prb}%
  \BibitemOpen
  \bibfield  {author} {\bibinfo {author} {\bibfnamefont {J.~K.}\ \bibnamefont
  {Block}}\ and\ \bibinfo {author} {\bibfnamefont {G.~M.}\ \bibnamefont
  {Bruun}},\ }\bibfield  {title} {\enquote {\bibinfo {title} {Properties of the
  density-wave phase of a two-dimensional dipolar fermi gas},}\ }\href
  {\doibase 10.1103/PhysRevB.90.155102} {\bibfield  {journal} {\bibinfo
  {journal} {Phys. Rev. B}\ }\textbf {\bibinfo {volume} {90}},\ \bibinfo
  {pages} {155102} (\bibinfo {year} {2014})}\BibitemShut {NoStop}%
\bibitem [{\citenamefont {Lee}\ \emph {et~al.}(2024)\citenamefont {Lee},
  \citenamefont {Baillie},\ and\ \citenamefont {Blakie}}]{lee_2024_PRA}%
  \BibitemOpen
  \bibfield  {author} {\bibinfo {author} {\bibfnamefont {Au-Chen}\ \bibnamefont
  {Lee}}, \bibinfo {author} {\bibfnamefont {D.}~\bibnamefont {Baillie}}, \ and\
  \bibinfo {author} {\bibfnamefont {P.~B.}\ \bibnamefont {Blakie}},\ }\bibfield
   {title} {\enquote {\bibinfo {title} {Excitations and phase ordering of the
  spin-stripe phase of a binary dipolar condensate},}\ }\href {\doibase
  10.1103/PhysRevA.109.023323} {\bibfield  {journal} {\bibinfo  {journal}
  {Phys. Rev. A}\ }\textbf {\bibinfo {volume} {109}},\ \bibinfo {pages}
  {023323} (\bibinfo {year} {2024})}\BibitemShut {NoStop}%
\bibitem [{\citenamefont {Zhang}\ \emph {et~al.}(2019)\citenamefont {Zhang},
  \citenamefont {Maucher},\ and\ \citenamefont {Pohl}}]{zhang19}%
  \BibitemOpen
  \bibfield  {author} {\bibinfo {author} {\bibfnamefont {Yong-Chang}\
  \bibnamefont {Zhang}}, \bibinfo {author} {\bibfnamefont {Fabian}\
  \bibnamefont {Maucher}}, \ and\ \bibinfo {author} {\bibfnamefont {Thomas}\
  \bibnamefont {Pohl}},\ }\bibfield  {title} {\enquote {\bibinfo {title}
  {Supersolidity around a critical point in dipolar bose-einstein
  condensates},}\ }\href {\doibase 10.1103/PhysRevLett.123.015301} {\bibfield
  {journal} {\bibinfo  {journal} {Phys. Rev. Lett.}\ }\textbf {\bibinfo
  {volume} {123}},\ \bibinfo {pages} {015301} (\bibinfo {year}
  {2019})}\BibitemShut {NoStop}%
\bibitem [{\citenamefont {Zhang}\ \emph {et~al.}(2021)\citenamefont {Zhang},
  \citenamefont {Pohl},\ and\ \citenamefont {Maucher}}]{zhang21}%
  \BibitemOpen
  \bibfield  {author} {\bibinfo {author} {\bibfnamefont {Yong-Chang}\
  \bibnamefont {Zhang}}, \bibinfo {author} {\bibfnamefont {Thomas}\
  \bibnamefont {Pohl}}, \ and\ \bibinfo {author} {\bibfnamefont {Fabian}\
  \bibnamefont {Maucher}},\ }\bibfield  {title} {\enquote {\bibinfo {title}
  {Phases of supersolids in confined dipolar bose-einstein condensates},}\
  }\href {\doibase 10.1103/PhysRevA.104.013310} {\bibfield  {journal} {\bibinfo
   {journal} {Phys. Rev. A}\ }\textbf {\bibinfo {volume} {104}},\ \bibinfo
  {pages} {013310} (\bibinfo {year} {2021})}\BibitemShut {NoStop}%
\bibitem [{\citenamefont {Zhang}\ \emph {et~al.}(2024)\citenamefont {Zhang},
  \citenamefont {Pohl},\ and\ \citenamefont {Maucher}}]{maucher24}%
  \BibitemOpen
  \bibfield  {author} {\bibinfo {author} {\bibfnamefont {Yong-Chang}\
  \bibnamefont {Zhang}}, \bibinfo {author} {\bibfnamefont {Thomas}\
  \bibnamefont {Pohl}}, \ and\ \bibinfo {author} {\bibfnamefont {Fabian}\
  \bibnamefont {Maucher}},\ }\bibfield  {title} {\enquote {\bibinfo {title}
  {Metastable patterns in one- and two-component dipolar bose-einstein
  condensates},}\ }\href {\doibase 10.1103/PhysRevResearch.6.023023} {\bibfield
   {journal} {\bibinfo  {journal} {Phys. Rev. Res.}\ }\textbf {\bibinfo
  {volume} {6}},\ \bibinfo {pages} {023023} (\bibinfo {year}
  {2024})}\BibitemShut {NoStop}%
\bibitem [{\citenamefont {Lima}\ \emph {et~al.}(2025)\citenamefont {Lima},
  \citenamefont {Grossklags}, \citenamefont {Zampronio}, \citenamefont
  {Cinti},\ and\ \citenamefont {Mendoza-Coto}}]{cinti2025}%
  \BibitemOpen
  \bibfield  {author} {\bibinfo {author} {\bibfnamefont {Daniel}\ \bibnamefont
  {Lima}}, \bibinfo {author} {\bibfnamefont {Matheus}\ \bibnamefont
  {Grossklags}}, \bibinfo {author} {\bibfnamefont {Vinicius}\ \bibnamefont
  {Zampronio}}, \bibinfo {author} {\bibfnamefont {Fabio}\ \bibnamefont
  {Cinti}}, \ and\ \bibinfo {author} {\bibfnamefont {Alejandro}\ \bibnamefont
  {Mendoza-Coto}},\ }\href {https://arxiv.org/abs/2501.09641} {\enquote
  {\bibinfo {title} {Supersolid dipolar phases in planar geometry: effects of
  tilted polarization},}\ } (\bibinfo {year} {2025}),\ \Eprint
  {http://arxiv.org/abs/2501.09641} {arXiv:2501.09641 [cond-mat.quant-gas]}
  \BibitemShut {NoStop}%
\bibitem [{\citenamefont {Zhang}\ \emph {et~al.}(2015)\citenamefont {Zhang},
  \citenamefont {Safavi-Naini}, \citenamefont {Rey},\ and\ \citenamefont
  {Capogrosso-Sansone}}]{amrey2015}%
  \BibitemOpen
  \bibfield  {author} {\bibinfo {author} {\bibfnamefont {C}~\bibnamefont
  {Zhang}}, \bibinfo {author} {\bibfnamefont {A}~\bibnamefont {Safavi-Naini}},
  \bibinfo {author} {\bibfnamefont {Ana~Maria}\ \bibnamefont {Rey}}, \ and\
  \bibinfo {author} {\bibfnamefont {B}~\bibnamefont {Capogrosso-Sansone}},\
  }\bibfield  {title} {\enquote {\bibinfo {title} {Equilibrium phases of tilted
  dipolar lattice bosons},}\ }\href {\doibase 10.1088/1367-2630/17/12/123014}
  {\bibfield  {journal} {\bibinfo  {journal} {New Journal of Physics}\ }\textbf
  {\bibinfo {volume} {17}},\ \bibinfo {pages} {123014} (\bibinfo {year}
  {2015})}\BibitemShut {NoStop}%
\bibitem [{\citenamefont {Bandyopadhyay}\ \emph {et~al.}(2019)\citenamefont
  {Bandyopadhyay}, \citenamefont {Bai}, \citenamefont {Pal}, \citenamefont
  {Suthar}, \citenamefont {Nath},\ and\ \citenamefont
  {Angom}}]{soumik:PRA:2019}%
  \BibitemOpen
  \bibfield  {author} {\bibinfo {author} {\bibfnamefont {Soumik}\ \bibnamefont
  {Bandyopadhyay}}, \bibinfo {author} {\bibfnamefont {Rukmani}\ \bibnamefont
  {Bai}}, \bibinfo {author} {\bibfnamefont {Sukla}\ \bibnamefont {Pal}},
  \bibinfo {author} {\bibfnamefont {K.}~\bibnamefont {Suthar}}, \bibinfo
  {author} {\bibfnamefont {Rejish}\ \bibnamefont {Nath}}, \ and\ \bibinfo
  {author} {\bibfnamefont {D.}~\bibnamefont {Angom}},\ }\bibfield  {title}
  {\enquote {\bibinfo {title} {Quantum phases of canted dipolar bosons in a
  two-dimensional square optical lattice},}\ }\href {\doibase
  10.1103/PhysRevA.100.053623} {\bibfield  {journal} {\bibinfo  {journal}
  {Phys. Rev. A}\ }\textbf {\bibinfo {volume} {100}},\ \bibinfo {pages}
  {053623} (\bibinfo {year} {2019})}\BibitemShut {NoStop}%
\bibitem [{\citenamefont {Wenzel}\ \emph {et~al.}(2017)\citenamefont {Wenzel},
  \citenamefont {B\"ottcher}, \citenamefont {Langen}, \citenamefont
  {Ferrier-Barbut},\ and\ \citenamefont {Pfau}}]{wenzel_2017:PRA}%
  \BibitemOpen
  \bibfield  {author} {\bibinfo {author} {\bibfnamefont {Matthias}\
  \bibnamefont {Wenzel}}, \bibinfo {author} {\bibfnamefont {Fabian}\
  \bibnamefont {B\"ottcher}}, \bibinfo {author} {\bibfnamefont {Tim}\
  \bibnamefont {Langen}}, \bibinfo {author} {\bibfnamefont {Igor}\ \bibnamefont
  {Ferrier-Barbut}}, \ and\ \bibinfo {author} {\bibfnamefont {Tilman}\
  \bibnamefont {Pfau}},\ }\bibfield  {title} {\enquote {\bibinfo {title}
  {Striped states in a many-body system of tilted dipoles},}\ }\href {\doibase
  10.1103/PhysRevA.96.053630} {\bibfield  {journal} {\bibinfo  {journal} {Phys.
  Rev. A}\ }\textbf {\bibinfo {volume} {96}},\ \bibinfo {pages} {053630}
  (\bibinfo {year} {2017})}\BibitemShut {NoStop}%
\bibitem [{\citenamefont {Macia}\ \emph
  {et~al.}(2012{\natexlab{a}})\citenamefont {Macia}, \citenamefont {Mazzanti},\
  and\ \citenamefont {Boronat}}]{macia12b}%
  \BibitemOpen
  \bibfield  {author} {\bibinfo {author} {\bibfnamefont {A.}~\bibnamefont
  {Macia}}, \bibinfo {author} {\bibfnamefont {F.}~\bibnamefont {Mazzanti}}, \
  and\ \bibinfo {author} {\bibfnamefont {J.}~\bibnamefont {Boronat}},\
  }\bibfield  {title} {\enquote {\bibinfo {title} {Ground state properties and
  excitation spectrum of a two dimensional gas of bosonic dipoles},}\ }\href
  {\doibase 10.1140/epjd/e2012-30455-y} {\bibfield  {journal} {\bibinfo
  {journal} {The European Physical Journal D}\ }\textbf {\bibinfo {volume}
  {66}},\ \bibinfo {pages} {301} (\bibinfo {year}
  {2012}{\natexlab{a}})}\BibitemShut {NoStop}%
\bibitem [{\citenamefont {Macia}\ \emph
  {et~al.}(2012{\natexlab{b}})\citenamefont {Macia}, \citenamefont {Hufnagl},
  \citenamefont {Mazzanti}, \citenamefont {Boronat},\ and\ \citenamefont
  {Zillich}}]{macia12}%
  \BibitemOpen
  \bibfield  {author} {\bibinfo {author} {\bibfnamefont {A.}~\bibnamefont
  {Macia}}, \bibinfo {author} {\bibfnamefont {D.}~\bibnamefont {Hufnagl}},
  \bibinfo {author} {\bibfnamefont {F.}~\bibnamefont {Mazzanti}}, \bibinfo
  {author} {\bibfnamefont {J.}~\bibnamefont {Boronat}}, \ and\ \bibinfo
  {author} {\bibfnamefont {R.~E.}\ \bibnamefont {Zillich}},\ }\bibfield
  {title} {\enquote {\bibinfo {title} {Excitations and stripe phase formation
  in a two-dimensional dipolar bose gas with tilted polarization},}\ }\href
  {\doibase 10.1103/PhysRevLett.109.235307} {\bibfield  {journal} {\bibinfo
  {journal} {Phys. Rev. Lett.}\ }\textbf {\bibinfo {volume} {109}},\ \bibinfo
  {pages} {235307} (\bibinfo {year} {2012}{\natexlab{b}})}\BibitemShut
  {NoStop}%
\bibitem [{\citenamefont {Macia}\ \emph {et~al.}(2014)\citenamefont {Macia},
  \citenamefont {Boronat},\ and\ \citenamefont {Mazzanti}}]{macia14}%
  \BibitemOpen
  \bibfield  {author} {\bibinfo {author} {\bibfnamefont {A.}~\bibnamefont
  {Macia}}, \bibinfo {author} {\bibfnamefont {J.}~\bibnamefont {Boronat}}, \
  and\ \bibinfo {author} {\bibfnamefont {F.}~\bibnamefont {Mazzanti}},\
  }\bibfield  {title} {\enquote {\bibinfo {title} {Phase diagram of dipolar
  bosons in two dimensions with tilted polarization},}\ }\href {\doibase
  10.1103/PhysRevA.90.061601} {\bibfield  {journal} {\bibinfo  {journal} {Phys.
  Rev. A}\ }\textbf {\bibinfo {volume} {90}},\ \bibinfo {pages} {061601}
  (\bibinfo {year} {2014})}\BibitemShut {NoStop}%
\bibitem [{\citenamefont {Bombin}\ \emph {et~al.}(2017)\citenamefont {Bombin},
  \citenamefont {Boronat},\ and\ \citenamefont {Mazzanti}}]{bombin17:PRL}%
  \BibitemOpen
  \bibfield  {author} {\bibinfo {author} {\bibfnamefont {R.}~\bibnamefont
  {Bombin}}, \bibinfo {author} {\bibfnamefont {J.}~\bibnamefont {Boronat}}, \
  and\ \bibinfo {author} {\bibfnamefont {F.}~\bibnamefont {Mazzanti}},\
  }\bibfield  {title} {\enquote {\bibinfo {title} {Dipolar bose supersolid
  stripes},}\ }\href {\doibase 10.1103/PhysRevLett.119.250402} {\bibfield
  {journal} {\bibinfo  {journal} {Phys. Rev. Lett.}\ }\textbf {\bibinfo
  {volume} {119}},\ \bibinfo {pages} {250402} (\bibinfo {year}
  {2017})}\BibitemShut {NoStop}%
\bibitem [{\citenamefont {Bomb\'{\i}n}\ \emph {et~al.}(2019)\citenamefont
  {Bomb\'{\i}n}, \citenamefont {Mazzanti},\ and\ \citenamefont
  {Boronat}}]{bombin19:PRA}%
  \BibitemOpen
  \bibfield  {author} {\bibinfo {author} {\bibfnamefont {Ra\'ul}\ \bibnamefont
  {Bomb\'{\i}n}}, \bibinfo {author} {\bibfnamefont {Ferran}\ \bibnamefont
  {Mazzanti}}, \ and\ \bibinfo {author} {\bibfnamefont {Jordi}\ \bibnamefont
  {Boronat}},\ }\bibfield  {title} {\enquote {\bibinfo {title}
  {Berezinskii-kosterlitz-thouless transition in two-dimensional dipolar
  stripes},}\ }\href {\doibase 10.1103/PhysRevA.100.063614} {\bibfield
  {journal} {\bibinfo  {journal} {Phys. Rev. A}\ }\textbf {\bibinfo {volume}
  {100}},\ \bibinfo {pages} {063614} (\bibinfo {year} {2019})}\BibitemShut
  {NoStop}%
\bibitem [{\citenamefont {Guijarro}\ \emph {et~al.}(2022)\citenamefont
  {Guijarro}, \citenamefont {Astrakharchik},\ and\ \citenamefont
  {Boronat}}]{guijarro22:PRL}%
  \BibitemOpen
  \bibfield  {author} {\bibinfo {author} {\bibfnamefont {G.}~\bibnamefont
  {Guijarro}}, \bibinfo {author} {\bibfnamefont {G.~E.}\ \bibnamefont
  {Astrakharchik}}, \ and\ \bibinfo {author} {\bibfnamefont {J.}~\bibnamefont
  {Boronat}},\ }\bibfield  {title} {\enquote {\bibinfo {title} {Ultradilute
  quantum liquid of dipolar atoms in a bilayer},}\ }\href {\doibase
  10.1103/PhysRevLett.128.063401} {\bibfield  {journal} {\bibinfo  {journal}
  {Phys. Rev. Lett.}\ }\textbf {\bibinfo {volume} {128}},\ \bibinfo {pages}
  {063401} (\bibinfo {year} {2022})}\BibitemShut {NoStop}%
\bibitem [{\citenamefont {Guijarro}\ \emph {et~al.}(2024)\citenamefont
  {Guijarro}, \citenamefont {Astrakharchik}, \citenamefont {Morigi},\ and\
  \citenamefont {Boronat}}]{guijarro2024}%
  \BibitemOpen
  \bibfield  {author} {\bibinfo {author} {\bibfnamefont {G.}~\bibnamefont
  {Guijarro}}, \bibinfo {author} {\bibfnamefont {G.~E.}\ \bibnamefont
  {Astrakharchik}}, \bibinfo {author} {\bibfnamefont {G.}~\bibnamefont
  {Morigi}}, \ and\ \bibinfo {author} {\bibfnamefont {J.}~\bibnamefont
  {Boronat}},\ }\bibfield  {title} {\enquote {\bibinfo {title} {Self-assembled
  chains and solids of dipolar atoms in a multilayer},}\ }\href {\doibase
  10.1103/PhysRevLett.133.233402} {\bibfield  {journal} {\bibinfo  {journal}
  {Phys. Rev. Lett.}\ }\textbf {\bibinfo {volume} {133}},\ \bibinfo {pages}
  {233402} (\bibinfo {year} {2024})}\BibitemShut {NoStop}%
\bibitem [{\citenamefont {Staudinger}\ \emph {et~al.}(2023)\citenamefont
  {Staudinger}, \citenamefont {Hufnagl}, \citenamefont {Mazzanti},\ and\
  \citenamefont {Zillich}}]{staudinger23:PRA}%
  \BibitemOpen
  \bibfield  {author} {\bibinfo {author} {\bibfnamefont {Clemens}\ \bibnamefont
  {Staudinger}}, \bibinfo {author} {\bibfnamefont {Diana}\ \bibnamefont
  {Hufnagl}}, \bibinfo {author} {\bibfnamefont {Ferran}\ \bibnamefont
  {Mazzanti}}, \ and\ \bibinfo {author} {\bibfnamefont {Robert~E.}\
  \bibnamefont {Zillich}},\ }\bibfield  {title} {\enquote {\bibinfo {title}
  {Striped dilute liquid of dipolar bosons in two dimensions},}\ }\href
  {\doibase 10.1103/PhysRevA.108.033303} {\bibfield  {journal} {\bibinfo
  {journal} {Phys. Rev. A}\ }\textbf {\bibinfo {volume} {108}},\ \bibinfo
  {pages} {033303} (\bibinfo {year} {2023})}\BibitemShut {NoStop}%
\bibitem [{\citenamefont {Delahaye}\ \emph {et~al.}(2019)\citenamefont
  {Delahaye}, \citenamefont {Chaimatanan},\ and\ \citenamefont
  {Mongeau}}]{delahaye:hal-01887543}%
  \BibitemOpen
  \bibfield  {author} {\bibinfo {author} {\bibfnamefont {Daniel}\ \bibnamefont
  {Delahaye}}, \bibinfo {author} {\bibfnamefont {Supatcha}\ \bibnamefont
  {Chaimatanan}}, \ and\ \bibinfo {author} {\bibfnamefont {Marcel}\
  \bibnamefont {Mongeau}},\ }\bibfield  {title} {\enquote {\bibinfo {title}
  {{Simulated annealing: From basics to applications}},}\ }in\ \href {\doibase
  10.1007/978-3-319-91086-4\_1} {\emph {\bibinfo {booktitle} {{Handbook of
  Metaheuristics}}}},\ \bibinfo {series} {International Series in Operations
  Research \& Management Science (ISOR)}, Vol.\ \bibinfo {volume} {272},\
  \bibinfo {editor} {edited by\ \bibinfo {editor} {\bibfnamefont {Michel}\
  \bibnamefont {Gendreau}}\ and\ \bibinfo {editor} {\bibfnamefont {Jean-Yves}\
  \bibnamefont {Potvin}}}\ (\bibinfo  {publisher} {{Springer}},\ \bibinfo
  {year} {2019})\ pp.\ \bibinfo {pages} {1--35.ISBN
  978--3--319--91085--7}\BibitemShut {NoStop}%
\bibitem [{\citenamefont {Leggett}(1970)}]{leggett70:prl}%
  \BibitemOpen
  \bibfield  {author} {\bibinfo {author} {\bibfnamefont {A.~J.}\ \bibnamefont
  {Leggett}},\ }\bibfield  {title} {\enquote {\bibinfo {title} {Can a solid be
  "superfluid"?}}\ }\href {\doibase 10.1103/PhysRevLett.25.1543} {\bibfield
  {journal} {\bibinfo  {journal} {Phys. Rev. Lett.}\ }\textbf {\bibinfo
  {volume} {25}},\ \bibinfo {pages} {1543--1546} (\bibinfo {year}
  {1970})}\BibitemShut {NoStop}%
\bibitem [{\citenamefont {Leggett}(1998)}]{leggett98:jsph}%
  \BibitemOpen
  \bibfield  {author} {\bibinfo {author} {\bibfnamefont {A.~J.}\ \bibnamefont
  {Leggett}},\ }\bibfield  {title} {\enquote {\bibinfo {title} {On the
  superfluid fraction of an arbitrary many-body system at t=0},}\ }\href
  {\doibase 10.1023/B:JOSS.0000033170.38619.6c} {\bibfield  {journal} {\bibinfo
   {journal} {Journal of Statistical Physics}\ }\textbf {\bibinfo {volume}
  {93}},\ \bibinfo {pages} {927--941} (\bibinfo {year} {1998})}\BibitemShut
  {NoStop}%
\bibitem [{\citenamefont {P\'erez-Cruz}\ \emph {et~al.}(2025)\citenamefont
  {P\'erez-Cruz}, \citenamefont {Astrakharchik},\ and\ \citenamefont
  {Massignan}}]{perez25:pra}%
  \BibitemOpen
  \bibfield  {author} {\bibinfo {author} {\bibfnamefont {Daniel}\ \bibnamefont
  {P\'erez-Cruz}}, \bibinfo {author} {\bibfnamefont {Grigori~E.}\ \bibnamefont
  {Astrakharchik}}, \ and\ \bibinfo {author} {\bibfnamefont {Pietro}\
  \bibnamefont {Massignan}},\ }\bibfield  {title} {\enquote {\bibinfo {title}
  {Superfluid fraction of interacting bosonic gases},}\ }\href {\doibase
  10.1103/PhysRevA.111.L011302} {\bibfield  {journal} {\bibinfo  {journal}
  {Phys. Rev. A}\ }\textbf {\bibinfo {volume} {111}},\ \bibinfo {pages}
  {L011302} (\bibinfo {year} {2025})}\BibitemShut {NoStop}%
\bibitem [{\citenamefont {Kadau}\ \emph
  {et~al.}(2016{\natexlab{b}})\citenamefont {Kadau}, \citenamefont {Schmitt},
  \citenamefont {Wenzel}, \citenamefont {Wink}, \citenamefont {Maier},
  \citenamefont {Ferrier-Barbut},\ and\ \citenamefont {Pfau}}]{Kadau:2016cb}%
  \BibitemOpen
  \bibfield  {author} {\bibinfo {author} {\bibfnamefont {Holger}\ \bibnamefont
  {Kadau}}, \bibinfo {author} {\bibfnamefont {Matthias}\ \bibnamefont
  {Schmitt}}, \bibinfo {author} {\bibfnamefont {Matthias}\ \bibnamefont
  {Wenzel}}, \bibinfo {author} {\bibfnamefont {Clarissa}\ \bibnamefont {Wink}},
  \bibinfo {author} {\bibfnamefont {Thomas}\ \bibnamefont {Maier}}, \bibinfo
  {author} {\bibfnamefont {Igor}\ \bibnamefont {Ferrier-Barbut}}, \ and\
  \bibinfo {author} {\bibfnamefont {Tilman}\ \bibnamefont {Pfau}},\ }\bibfield
  {title} {\enquote {\bibinfo {title} {{Observing the Rosensweig instability of
  a quantum ferrofluid}},}\ }\href@noop {} {\bibfield  {journal} {\bibinfo
  {journal} {Nature}\ }\textbf {\bibinfo {volume} {530}},\ \bibinfo {pages}
  {194--197} (\bibinfo {year} {2016}{\natexlab{b}})}\BibitemShut {NoStop}%
\end{thebibliography}%

\end{document}